\documentclass{tlp}

\usepackage{latexsym}
\usepackage{epsfig}
\usepackage{url}
\usepackage{amsmath}

\newcounter{theocounter}
\newtheorem{theorem}[theocounter]{Theorem}
\newcounter{lemcounter}
\newtheorem{lemma}[lemcounter]{Lemma}
\newcounter{procounter}
\newtheorem{proposition}[procounter]{Proposition}
\newcounter{corcounter}
\newtheorem{corollary}[corcounter]{Corollary}

\title[Bounded {LTL} Model Checking with Stable Models]
{Bounded {LTL} Model Checking with\\
Stable Models\thanks{This is an extended version of a paper
titled
``Bounded {LTL} Model Checking with Stable Models''
presented at the
6th International Conference on
Logic Programming and Nonmonotonic Reasoning ({LPNMR}'2001),
Vienna, Austria, September 2001.
}
}

\author[Keijo Heljanko and Ilkka Niemel{\"a}]
{KEIJO HELJANKO\thanks{The financial support of
Academy of Finland (Projects  53695, 47754) and
Foundation of Technology (Tekniikan Edist{\"a}mis\-s{\"a}{\"a}ti{\"o}), Helsinki, Finland
are gratefully acknowledged.}
and
ILKKA NIEMEL{\"A}\thanks{The financial support of
Academy of Finland (Projects  53695, 47754) is gratefully acknowledged.}\\
Helsinki University of Technology\\
Department of Computer Science and Engineering\\
Laboratory for Theoretical Computer Science\\
P.O.~Box 5400, FIN-02015 HUT, Finland\\
\email{\{Keijo.Heljanko, Ilkka.Niemela\}@hut.fi}
}

\newcommand{\lparrow}{\leftarrow}
\newcommand{\lpnot}{\textrm{not}\;}

\def\seq#1{\langle #1 \rangle}
\def\bbbn{{\rm I\!N}}

\newcommand{\corr}[1]{ #1 }

\newcommand{\myprogram}{\Pi}

\newcommand{\aprogram}[2]{\Pi_{\mathrm{A}}(#1,#2)}

\newcommand{\mprogram}[2]{\Pi_{\mathrm{M}}(#1,#2)}
\newcommand{\dprogram}[2]{\Pi_{\mathrm{D}}(#1,#2)}
\newcommand{\sprogram}[2]{\Pi_{\mathrm{S}}(#1,#2)}
\newcommand{\iprogram}[2]{\Pi_{\mathrm{I}}(#1,#2)}
\newcommand{\lprogram}[2]{\Pi_{\mathrm{L}}(#1,#2)}
\newcommand{\ltlprogram}[2]{\Pi_{\mathrm{LTL}}(#1,#2)}

\newcommand{\reduct}[2]{#1^{#2}}

\newcommand{\smodel}[1]{\Delta_{#1}}

\newcommand{\Smodels}{\texttt{Smodels}}
\newcommand{\Smodelsversion}{\texttt{Smodels~2.26}}

\newcommand{\IFFt}{iff }
\newcommand{\dlv}{\texttt{\small DLV}}
\newcommand{\NuSMV}{\texttt{NuSMV}}
\newcommand{\NuSMVversion}{\texttt{NuSMV~2.1.0}}
\newcommand{\NuSMVBMC}{\texttt{NuSMV/BMC}}
\newcommand{\NuSMVBDD}{\texttt{NuSMV/BDD}}
\newcommand{\zChaff}{\texttt{zChaff}}
\newcommand{\zChaffversion}{\texttt{zChaff~2001.2.17}}
\newcommand{\boundsmodels}{\texttt{boundsmodels}}
\newcommand{\boundsmodelsversion}{\texttt{boundsmodels~1.0}}

\newcommand{\nexttime}[2]{#1({#2} +1)}
\newcommand{\nexttrans}[1]{\mathit{{#1}}(n+1)}
\newcommand{\looppoint}{nl(i)}

\newcommand{\newarrow}[1]{\overset{#1}{\rightarrow}}

\newtheorem{definition}{Definition}

\begin{document}
\maketitle

\begin{abstract}
In this paper bounded model checking of asynchronous concurrent systems
is introduced as a promising application area for answer set
programming.
As the model of asynchronous systems a generalisation of communicating
automata, 1-safe Petri nets, are used.  It is shown how a 1-safe Petri
net and a requirement on the behaviour of the net can be translated into a
logic program such that the bounded model checking problem for the net
can be solved by computing stable models of the corresponding program.
The use of the stable model semantics leads to compact encodings of
bounded reachability and deadlock detection tasks as well as
the more general problem of bounded model checking of
linear temporal logic.
Correctness proofs of the devised translations are given, and
some experimental results
using the translation and the \Smodels\ system are presented.
\end{abstract}


\section{Introduction}

Recently, a novel paradigm for applying declarative logic
programming techniques has been proposed. 
In this approach, called \emph{answer set programming}
(a term coined by Vladimir Lifschitz), a problem
is solved by devising a logic program such that models
of the program provide the answers to the
problem~\cite{Lifschitz99:iclp,MT99:slp,Niemela99:amai}.  
Much of this work has been based on the stable model
semantics~\cite{GL88} and there are
efficient systems
\dlv\ (\url{http://www.dbai.tuwien.ac.at/proj/dlv/}) and
\Smodels\ (\url{http://www.tcs.hut.fi/Software/smodels/}) 
for computing stable\linebreak
models of logic programs. 
Using such an answer set programming system a problem is solved by
writing a logic program whose stable models capture the solutions of the
problem and then employing the system to compute a solution, i.e., a stable
model.

In this paper we put forward symbolic model
checking~\cite{Burch92,CGP99:book} as a promising application area for
answer set programming systems. In particular, we demonstrate how
bounded model checking problems of asynchronous concurrent systems can
be reduced to computing stable models of logic programs.

Verification of asynchronous systems is typically done by enumerating
the reachable states of the system.
Tools based on this approach (with various
enhancements) include, e.g., the \textsc{Spin} system~\cite{Holzmann97},
which supports extended finite state machines communicating through FIFO
queues, and the Petri net model based \textsc{PROD} tool~\cite{PROD32}.
The main problem with enumerative model checkers is the
amount of memory needed for the set of reachable states.

Symbolic model checking is widely applied especially in hardware
verification. The main analysis technique is based on (ordered) binary
decision diagrams (BDDs). In many cases the set of reachable states
can be represented very compactly using a BDD encoding.
Although the approach has been successful,
there are difficulties in applying BDD-based techniques, in particular,
in areas outside hardware verification. The key problem is that some
Boolean functions do not have a compact representation as BDDs and the
size of the BDD representation of a Boolean function is very sensitive
to the variable ordering.
Bounded model checking~\cite{BiereCimattiClarkeZhu:TACAS1999} has been
proposed as a technique for overcoming the space problem by replacing
BDDs with satisfiability (SAT) checking techniques because typical SAT checkers use only
polynomial amount of memory. The idea is roughly the following.  Given a
sequential digital circuit, a (temporal) property to be verified, and a bound $n$,
the behaviour of a sequential circuit is unfolded up to $n$ steps as a
Boolean formula $S$ and the negation of the property to be verified is
represented as a Boolean formula $\overline{R}$. The translation to Boolean
formulas is done so that $S \land \overline{R}$ is satisfiable iff the system has a
behaviour violating the property of length at most~$n$.  
Hence, bounded
model checking provides directly interesting and practically relevant
benchmarks for any answer set programming system capable of handling
propositional satisfiability problems.

Until now bounded model checking has been applied to synchronous
hardware verification and little attention has been given to knowledge
representation issues such as developing concise and efficient logical
representation of system behaviour. 
In this work we study the knowledge representation problem 
and employ ideas used in reducing planning to stable model
computation~\cite{Niemela99:amai}. 
The aim is to develop techniques such that the behaviour of an
asynchronous concurrent system can be encoded compactly and the
inherent concurrency in the system could be exploited in model checking
the system. 
To illustrate the approach we use
a simple basic Petri net model of asynchronous systems,
1-safe Place/Transition nets (P/T nets), which is an interesting
generalisation of communicating automata~\cite{DeselReisig:PT}.
Thus properties of finite state systems composed of finite state
machine components can be verified using model checkers for 1-safe Petri nets.

The structure of the rest of the paper is the following. In the next
section we introduce Petri nets and the bounded model checking problem.
Then we develop a compact encoding of bounded
model checking as the problem of finding stable models of logic
programs. 
We first show how to treat reachability properties such as
deadlocks
and then demonstrate how to extend the approach to cope with properties
expressed in linear temporal logic (LTL). 
We discuss initial experimental results and end with some concluding
remarks.   

\section{Petri nets and bounded model checking}
\begin{figure}
\begin{center}
\epsfig{file=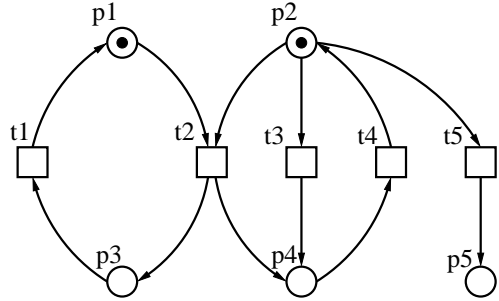,width=65mm}
\caption{Running Example}
\label{fig:running}
\end{center}
\end{figure}

There are several Petri net derived models presented in the literature.
We will use P/T-nets which are one of the simplest forms of Petri nets.
We will use as a running example the P/T-net presented in
Fig.~\ref{fig:running}. 

A triple $\seq{P, T, F}$ is a {\em net} if $P \cap T = \emptyset$ and
$F \subseteq (P \times T) \cup (T \times P)$. The elements of $P$
are called {\em places}, and the elements of $T$ {\em transitions}. Places
and transitions are also called {\em nodes}.
The places are represented in graphical notation by circles,
transitions by squares, and the {\em flow relation} $F$ by arcs.
We identify $F$ with its characteristic function on the set
$(P \times T) \cup (T \times P)$.
The {\em preset} of a node $x$,
denoted by ${}^{\bullet}x$, is the set
$\{y \in P \cup T \, \arrowvert \, F(y, x) = 1\}$.
In our running example, e.g.,  ${}^{\bullet}t2 = \{ p1, p2 \}$.
The {\em postset} of a node $x$,
denoted by $x^{\bullet}$, is the set
$\{y \in P \cup T \, \arrowvert \, F(x, y) = 1\}$.
Again in our running example
$p2^{\bullet} = \{ t2, t3, t5\}$.

A {\em marking} of a net $\seq{P, T, F}$ is a mapping $P \mapsto \bbbn$.
A marking $M$ is identified with the multi-set which contains $M(p)$
copies of $p$ for every $p \in P$.
A 4-tuple $\Sigma = \seq{P, T, F, M_0}$ is a {\em net system} (also
called a {\em P/T-net}) if 
$\seq{P, T, F}$ is a net and $M_0$ is a marking of $\seq{P, T, F}$
called the {\em initial marking}.
A marking is graphically denoted by a distribution of tokens on
the places of the net. In our running example in
Fig.~\ref{fig:running} the net has the initial marking
$M_0 = \{ p1, p2 \}$.

A marking $M$ enables a transition $t \in T$ if
$\forall p \in P: F(p,t) \leq M(p)$. If $t$ is enabled, it can {\em occur}
leading to a new marking (denoted $M \overset{t}{\rightarrow} M^{\prime}$),
where $M^{\prime}$ is
defined by $\forall p \in P: M^{\prime}(p) = M(p) - F(p,t) + F(t,p)$.
In the running example $t2$ is enabled in the initial
marking $M_0$, and thus $M_0 \overset{t2}{\rightarrow} M^{\prime}$, where
$M^{\prime} = \{ p3, p4 \}$.
A~marking $M$ is a {\em deadlock} if no transition $t \in T$ is enabled by $M$.
In our running example the marking $M = \{p1, p5\}$ is a deadlock.

A marking $M_n$ is {\em reachable} in $\Sigma$ if there is an {\em
execution}, i.e., a (possibly empty) sequence of transitions $t_0, t_1, \ldots, t_{n-1}$
and markings $M_1, M_2, \ldots, M_{n-1}$ such that:
$M_0 \overset{t_{0}}{\rightarrow} M_1 \overset{t_{1}}{\rightarrow} \ldots M_{n-1} \overset{t_{n-1}}{\rightarrow} M_{n}$.
A marking $M$ is reachable within a bound $n$, if
there is an execution with at most $n$ transitions, with which
$M$ is reachable.
The~net system may also have {\em infinite executions}, 
i.e., infinite sequences of transitions $t_0, t_1, \ldots$
and markings $M_1, M_2, \ldots$ such that:
$M_0 \overset{t_{0}}{\rightarrow} M_1 \overset{t_{1}}{\rightarrow} \ldots$.
The {\em maximal executions} of a net system are the
infinite executions of the net system together with the (finite) executions
leading to a deadlock marking.

A marking $M$ is 1-safe if $\forall p \in P: M(p) \leq 1$.
A P/T-net is 1-safe if all its reachable markings are 1-safe.
We will restrict ourselves to finite P/T-nets which are 1-safe,
and in which each transition has both nonempty pre- and
postsets.

Given a 1-safe P/T-net $\Sigma$, we say that a set
of transitions $S \subseteq T$ is {\em concurrently enabled} in the marking $M$, if
(i) all transitions $t \in S$ are enabled in $M$, and (ii)
for all pairs of transitions $t, t^{\prime} \in S$, such that $t \not = t^{\prime}$,
it holds that ${{}^{\bullet}t} \cap {{}^{\bullet}t^{\prime}} = \emptyset$.
If a set $S$ is concurrently enabled in the marking $M$, it can be fired
in a {\em step} (denoted $M \overset{S}{\rightarrow} M^{\prime}$),
where $M^{\prime}$ is the marking reached after firing all of the transitions
in the step $S$ in arbitrary order. It is easy to prove by using the 1-safeness
of the P/T-net $\Sigma$ that all possible interleavings of transitions in a step
$S$ are enabled in $M$, and that they all lead to the same final marking $M^{\prime}$.
In our running example in the marking $M^{\prime} = \{ p3, p4 \}$ the step
$\{ t1, t4 \}$ is enabled, and will lead back to the initial marking $M_0$. This
is denoted by $M^{\prime} \overset{\{ t1, t4 \}}{\rightarrow} M_0$.
Notice also that for any enabled transition, the singleton set containing only
that transition is
a step.

We say that a
marking $M_n$ is {\em reachable in step semantics} in a 1-safe P/T-net
if there is a {\em step execution}, i.e., a (possibly empty) sequences
$S_0, S_1, \ldots, S_{n-1}$ of steps
and $M_1, M_2, \ldots, M_{n-1}$ of markings such that:
$M_0~\overset{S_{0}}{\rightarrow}~M_1~\overset{S_{1}}{\rightarrow}~\ldots
M_{n-1}~\overset{S_{n-1}}{\rightarrow}~M_{n}$.
A~marking $M$ is reachable within a bound $n$ in the step semantics, if
there is a step execution with at most $n$ steps, with which
$M$ is reachable.
We will refer to the ``normal semantics''
as {\em interleaving semantics}.
The {\em infinite step executions} and {\em maximal step executions}
are defined in a similar way as in the interleaving case.

Note that if a marking is reachable in
$n$ transitions in the interleaving semantics, it is also reachable
in $n$ steps in the step semantics. However, the converse does not
necessarily hold.
We have, however, the following theorem which implicitly follows from the
results of~\cite{BD:TCS87}.
\begin{theorem}
For finite 1-safe P/T-nets the set of reachable markings in
the interleaving and step semantics coincide.
\end{theorem}

\paragraph{Linear temporal logic (LTL).}
The linear temporal logic LTL is one of the most widely used logics for
model checking reactive systems,
see e.g., \cite{CGP99:book}.
The basic idea is to specify properties that the system should have using
LTL.
A~model checker is then used to check whether all behaviours of the system
are models of the specification formula. If not, then the model checker
outputs a behaviour of the system which violates the given specification.

Given a finite set $AP$ of atomic propositions,
the syntax of LTL
is given by:
\[ \varphi ::=
p \in AP \,\, \arrowvert \,\,
\neg \varphi_{1} \,\, \arrowvert \,\,
\varphi_{1} \, \vee \, \varphi_{2} \,\, \arrowvert \,\,
\varphi_{1} \, \wedge \, \varphi_{2} \,\, \arrowvert \,\,
\varphi_{1} \, U \, \varphi_2 \,\, \arrowvert \,\,
\varphi_{1} \, R \, \varphi_2 \; . 
\]
Note that we do not define the often used
next-time operator $X \, \varphi_{1}$.
This is a commonly used tradeoff which in our case allows the combination of the step
semantics with LTL model checking.

We use $V = 2^{AP}$ as our alphabet.
We denote by $V^{+}$ all finite sequences over $V$ excluding the empty sequence,
and with $V^{\omega}$ all infinite sequences over $V$.
A word $w \in V^{+} \cup V^{\omega}$
is thus either a finite sequence $w = x_0\,x_1\,\ldots\, x_n$
or an infinite sequence $w = x_0\,x_1\,\ldots$, such that
$x_i \in V$ for all $i \geq 0$.
For a word $w$ we define $w_{(i)} = x_i$, and denote by
$w^{(i)}$ the suffix of $w$ starting at $x_i$.
When $w \in V^{+}$ we
define $|w|$ to be the length of the word $w$, and
in the case $w \in V^{\omega}$ we define $|w| = \omega$
where $\omega$ is greater than any natural number.

The relation $w \models \varphi$ is defined inductively as follows:
\begin{itemize}
\item $w \models p$ iff $p \in w_{(0)}$ for $p \in AP$,
\item $w \models \neg \varphi_{1}$ iff not $w \models \varphi_{1}$,
\item $w \models \varphi_{1} \, \vee \, \varphi_{2}$ iff
$w \models \varphi_{1}$ or $w \models \varphi_{2}$,
\item $w \models \varphi_{1} \, \wedge \, \varphi_{2}$ iff
$w \models \varphi_{1}$ and $w \models \varphi_{2}$,
\item $w \models \varphi_{1} \, U \, \varphi_2$ iff there exists
$0 \leq j  < |w|$, such that
$w^{(j)} \models \varphi_{2}$ and
for all $0 \leq i < j$, $w^{(i)} \models \varphi_{1}$,
\item $w \models \varphi_{1} \, R \, \varphi_2$ iff for all
$0 \leq j < |w|$,
if for every $0 \leq i < j$
$w^{(i)} \not \models \varphi_{1}$
then $w^{(j)} \models \varphi_{2}$ . 
\end{itemize}
We define some shorthand LTL formulas:
$\top \equiv p \, \vee \, \neg p$ for some arbitrary fixed $p \in AP$,
$\bot \equiv \neg \top$,
$\Diamond \, \varphi \equiv (\top \, U \, \varphi)$,
$\Box \, \varphi \equiv (\bot \, R \, \varphi)$,
and $\varphi_1 \rightarrow \varphi_2 \equiv \neg \varphi_1 \vee \varphi_2$.
The temporal operators are called: $U$ for ``until'', $R$ for ``release'',
$\Diamond$ for ``eventually'',
and $\Box$ for ``globally''.
Our definition of the semantics of LTL above is somewhat redundant.
This was done on purpose, as we often in this work use LTL formulas in
{\em positive normal form}, in which only a restricted use of
negations is allowed.
To be more specific,
an LTL formula is said to be in
positive normal form when all negations
in the formula appear directly before an atomic proposition.
A~formula can be put into positive normal form with
the following equivalences (and their duals):
$\neg \neg \varphi \equiv \varphi$,
$\neg (\varphi_1 \, \vee \, \varphi_2) \equiv \neg \varphi_1 \, \wedge \, \neg \varphi_2$,
and
$\neg (\varphi_1 \, U \, \varphi_2) \equiv \neg \varphi_1 \, R \, \neg \varphi_2$.
Note that converting a formula into positive normal form does not
involve a blowup.

Some examples of practical use of LTL formulas
are:
$\Box \neg(\mathit{cs}_1 \wedge \mathit{cs}_2)$ (it always holds that two processes are
not at the same time in a critical section),
$\Box (\mathit{req} \rightarrow \Diamond \mathit{ack})$
(it is always the case that a request is eventually followed by an acknowledgement),
and
$((\Box \Diamond \mathit{sch}_1) \wedge (\Box \Diamond \mathit{sch}_2)) \rightarrow (\Box (\mathit{tr}_1 \rightarrow \Diamond \mathit{cs}_1))$
(if both process 1 and 2 are scheduled infinitely often, then
always the entering of process 1 in the trying section
is followed by the process 1 eventually entering
the critical section).

Given a 1-safe P/T net $\Sigma$, we use a chosen subset of the places
as the atomic propositions $AP$.
A maximal (interleaving) execution
$M_0 \newarrow{t_0}M_1\newarrow{t_1}\ldots$ satisfies~$\varphi$ iff
the corresponding word
$w = (M_0 \cap AP), (M_1 \cap AP), \ldots$ satisfies $\varphi$. We say that
$\Sigma$ satisfies $\varphi$ iff every maximal execution starting
from the initial marking $M_0$ satisfies~$\varphi$.
Alternatively, $\Sigma$
does not satisfy $\varphi$ if there exists a maximal execution starting
from $M_0$ which satisfies $\neg \varphi$.
We call such an execution a {\em counterexample}.
Notice that we restrict ourselves to maximal executions
and thus our counterexamples are either infinite executions
or finite executions
leading to a deadlock (recall the
definition of maximal executions).

The temporal logic LTL can specify quite complex
properties of reactive systems.
In many cases it suffices to reason about much simpler temporal
properties. A~typical example is the reachability of a marking satisfying
some condition $C$ which in the LTL setting
corresponds to finding a counterexample for a formula $\Box \neg C$.
An important reachability based problem is deadlock detection.
\begin{definition}
{\bf(Deadlock detection)} Given a 1-safe P/T-net $\Sigma$, is there
a reachable marking $M$ which does not enable any transition
of $\Sigma$?
\end{definition}

Most analysis questions including deadlock detection and LTL model checking
are PSPACE-complete in the size of a 1-safe Petri net,
see e.g.,~\cite{Esparza:thumb}. In {\em bounded model checking} we
fix a bound $n$ and look for counterexamples which are
shorter than the given bound $n$.
For example,
in the case of {\em bounded deadlock detection} we look
for executions reaching a deadlock in at most $n$ transitions.
It is easy to show that, e.g.,
the bounded deadlock detection problem
is NP-complete (when the bound $n$ is given in unary coding).
This idea can also be applied to LTL model checking.
In~\cite{BiereCimattiClarkeZhu:TACAS1999}
{\em bounded LTL model checking} is introduced.
They also discuss how to ensure that a given bound $n$ is
sufficient to guarantee completeness.
Unfortunately, getting an exact bound is often computationally infeasible,
and easily obtainable upper bounds are too large.
In the case of 1-safe P/T-nets they are exponential in the number of places in the net.
Therefore the bounded
model checking results are usually not conclusive if a
counterexample is not found. Thus bounded model checking is at its
best in ``bug hunting'', and not as easily applicable in verifying
systems to be correct.

\newcommand{\smdel}{\Delta}
\newcommand{\smdelE}{\Delta_E}
\newcommand{\atoms}[1]{\textrm{Atoms}(#1)}
\newcommand{\execution}[3]{\sigma_{#1,#2}(#3)}
\newcommand{\LPSAT}{\mathrm{LP}}
\newcommand{\prop}{c}

\section{From bounded model checking to answer set programming}

In this section we show how to solve bounded LTL model checking problems
using answer set programming based on normal logic programs with the
stable model semantics. The basic idea is to reduce a bounded model
checking problem to a stable model computation task, i.e., to devise for
a P/T-net, a bound, and a temporal property to be checked a logic
program such that the stable models of the program correspond directly
to executions of the net within the bound violating the property.  Then
an implementation of the stable model semantics can be used to perform
bounded model checking tasks.  First we briefly review the stable model
semantics~\cite{GL88} and discuss a couple of useful shorthands to be
used in the encodings as well as the basis of an answer set programming
methodology with rules. Then we address the encoding of checking reachability
properties and finally extend the approach to handle full LTL model
checking.

\subsection{Stable model semantics}

For encoding bounded model checking problems we use
normal logic programs with
stable model semantics~\cite{GL88}.
A normal rule is of the form
\begin{equation}
a \lparrow b_1,\ldots,b_m,\lpnot c_{1},\ldots,\lpnot c_n
\label{eq:lprule}
\end{equation}
where each $a, b_i, c_j$ is a ground atom.
Models of a program are sets of ground atoms. 
A set of atoms $\smdel$ is said to satisfy an atom $a$ if $a \in \smdel$
and a negative literal $\lpnot {a}$ if $a \not\in \smdel$.
A rule $r$ of the form (\ref{eq:lprule}) is satisfied by $\smdel$ if the
head $a$ is satisfied whenever every body literal 
$b_1,\ldots,b_m,\lpnot c_{1},\ldots,\lpnot c_n$ is satisfied by
$\smdel$
and a program $\myprogram$ is
satisfied by $\smdel$ if each rule in $\myprogram$ is satisfied by
$\smdel$ (denoted $\smdel \models \myprogram$).

Stable models of a program are sets of ground atoms which satisfy all
the rules of the program and are justified by the rules. This is
captured using the concept of a \emph{reduct}. 
For a program $\myprogram$ and a set of atoms $\smdel$, the reduct 
$\reduct{\myprogram}{\smdel}$ is defined by
\[\reduct{\myprogram}{\smdel} = \{ a \lparrow b_1,\ldots,b_m \mid
\begin{array}[t]{@{}l}
a \lparrow b_1,\ldots,b_m,\lpnot c_{1},\ldots,\lpnot c_n \in \myprogram, \\
\{c_{1},\ldots,c_n\} \cap \smdel = \emptyset \}
\end{array}
\]
i.e., a reduct $\reduct{\myprogram}{\smdel}$ does not contain any
negative literals and, hence, has  
a unique subset minimal set of atoms satisfying it. 
\begin{definition}
A set of atoms $\smdel$ is a stable model of a program $\myprogram$ 
\IFFt $\smdel$ is
the unique minimal set of atoms satisfying $\reduct{\myprogram}{\smdel}$. 
\end{definition}
We employ three
extensions which can be seen as compact shorthands for normal rules.
We use \emph{integrity constraints}, i.e., rules 
\begin{equation}
\lparrow b_1,\ldots,b_m,\lpnot c_{1},\ldots,\lpnot c_n
\label{eq:ic}
\end{equation}
with
an empty head. Such a constraint can be taken as a
shorthand for a 
rule 
\[
f \lparrow \lpnot f, b_1,\ldots,b_m,\lpnot c_{1},\ldots,\lpnot c_n
\]
where $f$ is a new atom. 
Notice that a stable model $\smdel$ satisfies an integrity constraint
(\ref{eq:ic}) only if at least one of its body literals
is not satisfied by $\smdel$.

For expressing the choice whether to include an atom in a stable model
we use \emph{choice rules}. They are normal rules where
the head is in brackets with the idea that the head  can be 
included in a stable model only if the body holds but it can be left
out, too.  Such a construct can be
represented using normal rules by introducing a new atom. 
For example, the choice rule on the left corresponds to the two normal rules on
the right where $a'$ is a new atom. 
\[
\begin{array}[t]{l}
\{ a \} \lparrow b, \lpnot c
\end{array}
\hspace{3em}\leadsto\hspace{3em}
\begin{array}[t]{l}
a \lparrow \lpnot a', b, \lpnot c\\
a' \lparrow \lpnot a
\end{array}
\]
Finally, a compact encoding of \emph{conflicts} is needed, i.e., 
rules of the form
\begin{equation}
\lparrow 2 \{ a_1,\ldots,a_n \}
\label{eq:conflict}
\end{equation}
saying that a stable model cannot contain any two
atoms out of a 
set of atoms $\{ a_1,\ldots,a_n \}$. 
Such a rule can be expressed, e.g., by adding a rule 
$f \lparrow \lpnot f, a_i, a_j$, where $f$ is a new atom,
for each pair $a_i, a_j$ from $\{
a_1,\ldots,a_n \}$, i.e., using ${\cal O}(n^2)$ rules.
Choice and conflict rules are simple cases of cardinality constraint
rules~\cite{NS2000:lbai}.
The \Smodels\ system
provides an implementation for cardinality constraint rules and includes
primitives supporting directly such constraints without translating them
first to corresponding normal rules. 

A straightforward method of using logic program rules  for 
answer set programming can be based on a \emph{generate and test} idea. 
A set of rules plays the role of a generator capturing stable models
corresponding to all candidate solutions and another set of
rules, testers, eliminate the non-valid ones.
A systematic way of using this method can be based on some simple
modularity properties of stable model semantics which are given below as
propositions where the first two are straightforward consequences of the
splitting theorem~\cite{LT94:iclp}. 

The propositions play an important role in
proving the correctness of our logic program encodings.
The first one says that if rules defining new atoms are added, then
a stable model of the original program can be obtained directly from
a stable model of the extended program. 
Often a tester is encoded using a stratified set of rules and an
integrity constraint. The next two propositions show that this does not
introduce new stable models but extends the original ones and possibly eliminates
some of them. 
\begin{proposition}
Let $\myprogram_1$ and $\myprogram_2$ be programs such that the atoms in
the heads of the rules in $\myprogram_2$ do not occur in $\myprogram_1$. 
Then for every stable model $\smdel$ of $\myprogram_1 \cup
\myprogram_2$, $\smdel \cap \atoms{\myprogram_1}$ is a 
stable model of $\myprogram_1$ where $\atoms{\myprogram_1}$ denotes the
set of atoms appearing in $\myprogram_1$. 
\label{pro:projection}
\end{proposition}
\begin{proposition}
Let $\myprogram_1$ be a program and $\myprogram_2$ a stratified program
such that the atoms in the heads of the rules in $\myprogram_2$ do not
occur in $\myprogram_1$.  
Then for every stable model $\smdel_1$ of $\myprogram_1$
there is a unique stable model $\smdel$ of $\myprogram_1  \cup
\myprogram_2$ such that $\smdel_1 = \smdel \cap \atoms{\myprogram_1}$.
\label{pro:strat:extension}
\end{proposition}
\begin{proposition}
Let $\myprogram$ be a program. 
Then $\smdel$ is a stable model of $\myprogram$ and 
satisfies an integrity constraint $ic$ (\ref{eq:ic}) iff 
 $\smdel$ is a stable model of $\myprogram \cup \{ic\}$. 
\label{pro:ic:extension}
\end{proposition}

\subsection{Reachability checking}

Now we devise a method for translating bounded reachability 
problems of 1-safe P/T-nets to tasks of finding stable models.
Consider a net $N = \seq{P, T, F}$ and a step bound $n \geq 1$. 
We construct a logic program $\aprogram{N}{n}$, which captures the
possible executions of $N$ up to $n$ steps,
as follows.
\begin{itemize}
\item For each place $p \in P$, include a choice rule
\begin{equation}
\{p(0)\} \lparrow \; .
\label{eq:initialmarking}
\end{equation}
\item For each transition $t \in T$, and for all $i = 0,1,\ldots,n-1$, include
a rule 
\begin{equation}
\{t(i)\} \lparrow p_1(i), \ldots, p_l(i)
\label{eq:tchoices}
\end{equation}
where $\{p_1, \ldots, p_l\}$ is the preset of $t$. 
Hence, a stable model can contain a transition instance in step $i$ only
if its preset holds at step $i$. 

\item For each place $p \in P$, for each transition $t$ in the preset of $p$,
and for all $i = 0,1,\ldots,n-1$, include a rule 
\begin{equation}
p(i+1)\lparrow t(i) \; .
\label{eq:effects}
\end{equation}
These say that $p$ holds in the next step if at least one of its preset
transitions is in the current step. 

\item For each place $p \in P$, and for all $i = 0,1,\ldots,n-1$,
if the cardinality of the postset $\{ t_1, \ldots, t_l\}$ of $p$
is a least 2, include a rule 
\begin{equation}
\lparrow 2 \{ t_1(i), \ldots, t_l(i) \} \; .
\label{eq:conflicts}
\end{equation}
This rule states that at most one of the transitions
that are in conflict w.r.t.\ $p$ can occur. 

\item For each place $p$, and for all $i = 0,1,\ldots,n-1$,
\begin{equation}
p(i+1) \lparrow p(i), \lpnot t_1(i),\ldots,\lpnot t_l(i)
\label{eq:frame}
\end{equation}
where $\{ t_1,\ldots,t_l\}$ is the set of transitions having $p$ in
their preset. This is the \emph{frame axiom} for $p$ stating that 
$p$ continues to hold if no transition using it occurs.
\item Disallow execution of transitions followed by idling.
For all $i = 0,1,\ldots,n-1$, include rules
\begin{equation}
\mathit{idle}(i) \lparrow \lpnot t_1(i), \ldots, \lpnot t_k(i)
\hspace{1em}
\lparrow \mathit{idle}(i+1), \lpnot \mathit{idle}(i)
\label{rule:idlestart}
\end{equation}
where $\{ t_1, \ldots, t_k \} = T$, i.e., the set of all transitions.
These rules force all idling to happen at the beginning, followed
by non-idling time-steps (if any).
\end{itemize}
As an example consider net $N$ in Fig.~\ref{fig:running} for which program
$\aprogram{N}{n}$ is given in Fig.~\ref{fig:rprogram}. 
\begin{figure}[thb]
\centering
$
\hfill
\begin{array}[t]{l}
\{p1(0)\} \lparrow\\ 
\{p2(0)\} \lparrow\\ 
\{p3(0)\} \lparrow\\ 
\{p4(0)\} \lparrow\\ 
\{p5(0)\} \lparrow\\
\{ t1(i) \} \lparrow p3(i)\\ 
\{ t2(i) \} \lparrow p1(i), p2(i)\\ 
\{ t3(i) \} \lparrow p2(i)\\ 
\{ t4(i) \} \lparrow p4(i)\\ 
\{ t5(i) \} \lparrow p2(i)\\
p1(i+1) \lparrow t1(i)\\ 
p2(i+1) \lparrow t4(i)\\
p3(i+1) \lparrow t2(i)\\ 
p4(i+1) \lparrow t2(i)\\ 
p4(i+1) \lparrow t3(i)\\ 
p5(i+1) \lparrow t5(i)\\
\end{array}
\hfill
\begin{array}[t]{l}
\lparrow 2 \{  t2(i), t3(i), t5(i)\}\\
p1(i+1) \lparrow p1(i), \lpnot t2(i)\\
p2(i+1) \lparrow \begin{array}[t]{@{}l}
p2(i), \lpnot t2(i), \lpnot t3(i),\\
\lpnot t5(i)
\end{array}\\
p3(i+1) \lparrow p3(i), \lpnot t1(i)\\
p4(i+1) \lparrow p4(i), \lpnot t4(i)\\
p5(i+1) \lparrow p5(i)\\
\mathit{idle}(i) \lparrow \begin{array}[t]{@{}l}\lpnot t1(i), \lpnot t2(i), \lpnot t3(i),\\
\lpnot t4(i),
\lpnot t5(i)
\end{array}\\
\lparrow \mathit{idle}(i+1), \lpnot \mathit{idle}(i)\\
\\
\mbox{  where } i = 0,1,\ldots n-1\\
\end{array}
\hfill
$
\caption{Program $\aprogram{N}{n}$ }
\label{fig:rprogram}
\end{figure}

In $\aprogram{N}{n}$ the initial marking is not constrained. 
Next we show how to limit markings using rules, i.e., how to
construct a set of rules $\mprogram{C}{i}$ that eliminates all stable
models which do not satisfy a given Boolean expression $C$ of marking
conditions at step $i$. The set $\mprogram{C}{i}$ includes the rule 
$\lparrow \lpnot c(i)$ and a set of rules defining $c(i)$ by a systematic
translation of the condition $C$ at step $i$ as explained next. 
A Boolean expression $C$ can be encoded with
rules by introducing for each non-atomic subexpression of $C$ a new atom
together with rules capturing the conditions under which the
subexpression is satisfied in the following way~\cite{NS2000:lbai}. 
Given a Boolean expression $C$ with connectives $\neg, \lor,
\land$, every subexpression of $C$ of the form 
$\neg \phi$ is mapped to a rule $\prop_{\neg\phi} \lparrow
\lpnot\prop_\phi$; 
a subexpression $\phi\land\psi$  is mapped to 
$\prop_{\phi\land\psi} \lparrow  \prop_\phi,\prop_\psi$ 
and $\phi\lor\psi$ to the two rules 
$\prop_{\phi\lor\psi} \lparrow  \prop_\phi$ and 
$\prop_{\phi\lor\psi} \lparrow  \prop_\psi$ where 
$\prop_\psi, \prop_\phi$ are new atoms introduced for the non-atomic
subexpressions. These are not needed for the atomic ones, i.e., $\prop_a
= a$ for an atom $a$. The conditions for a step $i$ are then obtained by
indexing all atoms with $i$.

The encoding of marking conditions is illustrated 
by considering a condition $C: p_1 \land (\neg p_2 \lor p_3)$ 
saying that $p_1 \in M$ and ($p_2 \not \in M$ or $p_3 \in M$)
and a step $i$. Now the set of rules
$\mprogram{C}{i}$ is
\[
\begin{array}[t]{ll}
\lparrow \lpnot c(i) \\
c(i) \lparrow p_1(i), c_{\neg p_2 \lor p_3 } (i)
\end{array}
\hspace{2em}
\begin{array}[t]{l}
 c_{\neg p_2 \lor p_3} (i)  \lparrow c_{\neg p_2}(i) \\
 c_{\neg p_2 \lor p_3}(i) \lparrow p_3(i) 
\end{array}
\hspace{2em}
\begin{array}[t]{l}
 c_{\neg p_2} (i)  \lparrow \lpnot p_2(i) \\
\end{array}
\]

Our approach can 
solve a reachability problem for a set of initial markings 
given by a condition $C_0$  where the markings to be reached are
specified by another condition $C$.  
\begin{theorem}
\label{th:reachability}
  Let $N = \seq{P, T, F}$ be a 1-safe P/T-net for all initial markings
  satisfying a condition $C_0$.
  Net $N$ has an initial marking satisfying $C_0$
  such that a marking satisfying a condition $C$ is reachable in at most $n$ steps iff
  $\mprogram{C_0}{0} \cup \aprogram{N}{n} \cup \mprogram{C}{n}$ has a
  stable model. 
\end{theorem}
\begin{proof}
See Appendix~\ref{app:reachability}.
\end{proof}

The deadlock detection problem is now just a special case of a
reachability property where the rules $\mprogram{C}{n}$ exclude markings
with some transition enabled. This set of rules is denoted by 
$\dprogram{N}{n}$ and it consists of 
the rule $\lparrow \mathit{live}$ and the program $\lprogram{N}{n}$ 
which includes for each transition
$t \in T$ and its preset
$\{ p_1,\ldots,p_l \}$, 
a rule 
\begin{equation}
\mathit{live} \lparrow p_1(n),\ldots,p_l(n) \; .
\label{eq:live}
\end{equation}
For our running example, the rules $\lprogram{N}{n}$
are 
\[
\mathit{live} \lparrow p3(n) \hspace{2em}
\mathit{live} \lparrow p1(n), p2(n) \hspace{2em}
\hspace{2em}
\mathit{live} \lparrow p2(n) \hspace{2em}
\mathit{live} \lparrow p4(n) \; .
\]

\subsection{Bounded LTL model checking}

Our strategy for finding counterexamples for LTL formula $\varphi$
(i.e., executions satisfying $\neg \varphi$) is almost the same
as in \cite{BiereCimattiClarkeZhu:TACAS1999}. The main difference
is that we allow the system under model checking to have reachable
deadlocks, while their translation does not allow this.
This is also a difference to our previous work~\cite{HelNie:LPNMR2001}.

\begin{figure}[thb]
\begin{center}
\input{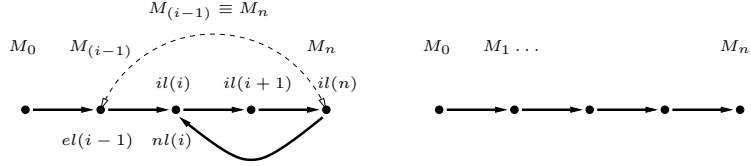}
\caption{Two counterexample possibilities}
\label{fig:cex}
\end{center}
\end{figure}

Our counterexamples have two basic shapes.
On the left in Fig.~\ref{fig:cex} is
a {\em loop counterexample}, and on the right is a {\em counterexample without loop}.
Loop counterexamples specify an infinite execution, while counterexamples
without a loop specify a~finite execution.
The arcs of the figure denote the ``next state'' of each state. Notice in the loop
counterexample that if $M_{(i-1)}$ is equivalent to the last state $M_n$, the state
$M_i$ is the ``next state'' of $M_n$.
The counterexamples without loop can additionally be divided into {\em deadlock executions}
(ending in a deadlock state), and {\em non-maximal executions} (ending in a state
which is not a deadlock).

In the case of non-maximal executions our encoding is a cautious one,
and we will find counterexamples which exist, no matter how the
non-maximal execution is extended into a maximal one.
(Recall that we have defined the semantics of LTL over maximal executions of the
net system.) Finding non-maximal counterexample executions is in fact
only an optimisation. It was introduced in~\cite{BiereCimattiClarkeZhu:TACAS1999},
and allows some counterexamples to be found with smaller
bounds than would otherwise be possible.

In the encoding we use the auxiliary atoms 
$\mathit{el}(i), \mathit{le}, \mathit{nl}(i), \mathit{il}(i)$ with
following intuition (see Fig.~\ref{fig:cex} for an example). 
The $\mathit{el}(i)$ atom is in a stable model for the state $i$ that is
equivalent with the last state $n$ and $\mathit{le}$ is in the model if
a loop exists, i.e., some $\mathit{el}(i)$ is in the model. 
The $\mathit{nl}(i)$ atom is in a model for the
``next state'' $i$ of the last state, while $\mathit{il}(i)$ is in the
model for all states $i$ in the loop. 

Given an LTL formula $f$ in positive normal form\footnote{
Using the positive normal form is required to handle non-maximal
counterexample executions, for which the duality
$f_1 \, R \, f_2 \equiv \neg (\neg f_1 \, U \, \neg f_2)$ can not be used,
see~\cite{BiereCimattiClarkeZhu:TACAS1999}.
}
(when the formula to be model checked is $\varphi$, the formula $f$ is
equivalent to $\neg \varphi$ with negations pushed in), and a bound $n
\geq 1$ we construct a program $\ltlprogram{f}{n}$ as follows.

\begin{itemize}
\item Guess which state is equivalent to the last (if any). For all $0 \leq i \leq n-1$ add rule
\begin{equation}
\{ \mathit{el}(i) \} \lparrow \; .\label{rule:guessloop}
\end{equation}
\item Disallow guessing two or more. (Guessing none is allowed though.) Add rule
\begin{equation}
\lparrow 2 \{ \mathit{el}(0), \mathit{el}(1), \ldots, \mathit{el}(n-1) \} \; .\label{rule:guesscheck}
\end{equation}
\item Check that the guess is correct. For all $0 \leq i \leq n-1$, $p \in P$ include rules
\begin{equation}
  \lparrow \mathit{el}(i), p(i), \lpnot p(n) \hspace{2em}
  \lparrow \mathit{el}(i), p(n), \lpnot p(i) \; .\label{rule:checkloop}
\end{equation}
\item Specify auxiliary loop related atoms.
For all $0 \leq i \leq n-1$, include rules
\begin{equation}
\mathit{le} \lparrow \mathit{el}(i) \hspace{1em}
\mathit{nl}(i+1) \lparrow \mathit{el}(i) \hspace{1em}
\mathit{il}(i+1) \lparrow \mathit{el}(i) \hspace{1em}
\mathit{il}(i+1) \lparrow \mathit{il}(i) \; .\label{rule:loopstuff}
\end{equation}
\item Require that if a loop exists, the last step contains a transition
to disallow looping by idling. Add the rule
\begin{equation}
\lparrow \mathit{le}, \mathit{idle}(n-1) \; .\label{rule:noidleloop}
\end{equation}
\item Allow at most one visible transition in a step to eliminate 
steps which cannot be interleaved to yield a counterexample.
For all $0 \leq i \leq n-1$, add rule
\begin{equation}
\lparrow 2 \{ t_1(i), \ldots, t_k(i)\} \label{rule:onevisible}
\end{equation}
where $\{ t_1, \ldots, t_k \}$ is the set of {\em visible
  transitions}, i.e., 
the transitions whose firing changes the marking of a place $p$ appearing in the formula $f$.
More formally, a transition $t \in T$ is visible, if there exists a place $p \in AP$
such that $F(t,p) - F(p,t) \not = 0$.
\end{itemize}
\begin{figure}
\begin{center}
\begin{tabular}{clcl}
\hline
Formula type&Translation&Formula type&Translation\\
\hline
\begin{oldtabular}{l}
${p, \mbox{ for } p \in AP}$
\end{oldtabular} &
\begin{oldtabular}{l}
$f(i) \lparrow p(i)$
\end{oldtabular} &
\begin{oldtabular}{l}
${\neg p, \mbox{ for } p \in AP}$
\end{oldtabular} &
\begin{oldtabular}{l}
$f(i) \lparrow \lpnot p(i)$
\end{oldtabular}\\
\hline
\begin{oldtabular}{l}
${f_1 \, \vee \, f_2}$
\end{oldtabular} &
\begin{oldtabular}{l}
$f(i) \lparrow f_1(i)$\\
$f(i) \lparrow f_2(i)$\\
\end{oldtabular} &
\begin{oldtabular}{l}
${f_1 \, \wedge \, f_2}$
\end{oldtabular} &
\begin{oldtabular}{l}
$f(i) \lparrow f_1(i), f_2(i)$
\end{oldtabular}\\
\hline
${f_1 \, U \, f_2}$ &
\begin{oldtabular}{l}
$f(i) \lparrow f_2(i)$\\
$f(i) \lparrow f_1(i), \nexttime{f}{i}$\\
$\nexttrans{f} \lparrow \looppoint, f(i)$
\end{oldtabular} &
${f_1 \, R \, f_2}$ &
\begin{oldtabular}{l}
$f(i) \lparrow f_2(i), f_1(i)$\\
$f(i) \lparrow f_2(i), \nexttime{f}{i}$\\
$\nexttrans{f} \lparrow \looppoint, f(i)$\\
$f(n+1) \lparrow le, \lpnot \mathit{c}(f)$\\
$\mathit{c}(f) \lparrow \mathit{il}(i), \lpnot f_2(i)$\\
$f(n) \lparrow f_2(n), \lpnot \mathit{live}$
\end{oldtabular}\\
\hline
\end{tabular}
\caption{Translation of an LTL formula $f$}\label{fig:trans}
\end{center}
\end{figure}

We recursively translate the formula $f$ by first translating its subformulas,
and then  $f$ as follows. For all $0 \leq i \leq n$, add the rules given
by Fig.~\ref{fig:trans}.\footnote{An equivalence explaining
the release translation:
$f_1 \, R \, f_2 \equiv (f_2 \, U \, (f_1 \, \wedge \, f_2 )) \, \vee \, (\Box f_2)$.}
Finally we require that the top level formula $f$ should hold in the initial marking
\begin{equation}
\lparrow \lpnot f(0) \; .\label{eq:topok}
\end{equation}
With this program $\ltlprogram{f}{n}$ we get our main result.
\begin{theorem}
  \label{th:ltl}
  Let $f$ be an LTL formula in positive normal form and
  $N = \seq{P, T, F}$ be a 1-safe P/T-net for all initial markings
  satisfying a condition $C_0$.
  If $\mprogram{C_0}{0} \cup \aprogram{N}{n} \cup \lprogram{N}{n} \cup \ltlprogram{f}{n}$
  has a stable model, then there is a maximal execution of
  $N$ from an initial marking satisfying $C_0$ which satisfies~$f$.
\end{theorem}
\begin{proof}
See Appendix~\ref{app:ltl1}.
\end{proof}

We also have the following completeness result for our translation.
First we define the notion of a {\em looping execution}.
A finite execution
$M_0 \overset{t_{0}}{\rightarrow} M_1 \overset{t_{1}}{\rightarrow} \ldots M_{n-1} \overset{t_{n-1}}{\rightarrow} M_{n}$ is a looping execution, if $n \geq 1$ and there exists an index $l < n$ such that $M_l = M_n$.
A looping execution together with the index $l$
is a finite witness to the existence of the corresponding (infinite) maximal execution $\sigma$ of the
net system $N$ which visits the sequence of states $M_0, M_1, \ldots, M_l, M_{l+1}, \ldots, M_k, M_{l+1}, \ldots, M_k, \ldots$.
\begin{theorem}
  \label{th:compl}
  Let $f$ be an LTL formula in positive normal form and
  $N = \seq{P, T, F}$ be a 1-safe P/T-net for all initial markings
  satisfying a condition $C_0$.
  If $N$ has a looping or deadlock execution of at most length $n$
  starting from an initial marking satisfying $C_0$ such that some corresponding
  maximal execution $\sigma$ satisfies $f$, then
  $\mprogram{C_0}{0} \cup \aprogram{N}{n} \cup \lprogram{N}{n} \cup \ltlprogram{f}{n}$
  has a stable model.
\end{theorem}
\begin{proof}
See Appendix~\ref{app:compl}.
\end{proof}

The size of the program in Theorem~\ref{th:ltl} is linear in the size of the
net
and formula, i.e., ${\cal O}((\arrowvert P \arrowvert + \arrowvert T \arrowvert + \arrowvert F \arrowvert +
\arrowvert f \arrowvert) \cdot n)$.
The semantics of LTL is defined over interleaving executions.
A novelty of the translation is that it allows concurrency between
invisible transitions.

We could simplify the LTL translation presented above
in following ways.
Firstly, if the net system is known to be deadlock free,
the release translation in Fig.~\ref{fig:trans} can be simplified by
removing the rule
$$f(n) \lparrow f_2(n), \lpnot \mathit{live},$$
and also the (now unnecessary) subprogram $\lprogram{N}{n}$.

Secondly, if we remove
the possibility of obtaining non-maximal counterexample
executions, the release translation can be removed
fully by using the equivalence
$\varphi_{1} \, R \, \varphi_2 \equiv \neg \, (\neg \, \varphi_{1} \, U \, \neg \, \varphi_2)$
and adding (the obvious) translation for negation.
This can not be done when non-maximal counterexamples are
used, because the equivalence does not hold in that
case.
As an example, one can not deduce from the fact that
$\neg \Diamond \, \neg \, \varphi$ holds for a non-maximal execution $\sigma$
that $\Box \, \varphi$ holds for any maximal
execution $\sigma'$ such that $\sigma$ is a prefix of $\sigma'$.
The non-maximal counterexample executions are
quite valuable in practice, as using them violations to safety
properties can be found with smaller bounds.
Therefore we chose to use a more complicated
translation for release.

\paragraph{Forcing interleaving semantics.}
We can create the interleaving semantics versions
of bounded model checking problems by adding
a set of rules $\iprogram{N}{n}$.
It includes for each time step
$0 \leq i \leq n-1$ a rule
\begin{equation}
\lparrow 2 \{ t_1(i), \ldots, t_m(i) \}
\label{eq:nostep}
\end{equation}
where $\{ t_1, \ldots, t_m \}$ is the set of all transitions.
These rules eliminate all stable models having more than one
transition firing in a step.
\begin{corollary}
  Let $\sprogram{N}{n}$ be a program solving a bounded
  model checking problem in the step semantics using any of
  the translations above. Then the program
  $\sprogram{N}{n} \cup \iprogram{N}{n}$ solves
  the same problem in the interleaving semantics.
\end{corollary}

\subsection{Relation to previous work}

Logic programming techniques have been used to model checking
branching time modal logics like the modal mu-calculus and CTL where model
checking can be reduced to solving equations with least and greatest
fixed points. A state of the art example of this approach is the XMC
system~\cite{XMC00:cav} which has been extended to handle also
linear temporal logic LTL using the standard tableau style
approach~\cite{PR00:tapd}. This method has the disadvantage that the
size of the resulting tableau can be exponential w.r.t. the size of the
temporal formula to be checked. The exponential worst case space
complexity, which is present in typical LTL model checkers,  is avoided
in bounded model checking where the space complexity remains polynomial
also w.r.t. the temporal formula. 

In previous work on bounded model checking little attention has been
given to the knowledge representation problem of encoding succinctly the
unfolded behavior and the temporal property. 
We address this problem and develop an encoding of 
the behavior of an asynchronous system which is linear in the size of
the system description (Petri net) and the formula as well as in the
number of steps. 

Our approach extends the previous work in several respects. 
Earlier research has been based on the interleaving semantics. 
Our work allows the use of the step semantics which enables
the exploitation of the inherent concurrency of the system in model
checking. The standard approach~\cite{BiereCimattiClarkeZhu:TACAS1999} 
assumes that the system to be model-checked is deadlock-free while we
can do LTL model checking for systems with reachable deadlocks.

We develop a more compact encoding of bounded LTL model checking. 
Our encoding is linear in the size of the net, the formula and the
bound. 
In~\cite{BiereCimattiClarkeZhu:TACAS1999} the encoding 
is superlinear in the size of the formula. The
paper provides no upper bound on the size 
w.r.t. the formula but states that it is polynomial in the size of the
formula if common subexpressions are shared and quadratic in the bound. 
These same observations can also be made of the optimised version of
the translation presented in~\cite{NuSMV:BMC}.
The compactness of our encoding is due to the
fact that the stable model semantics supports least fixed point
evaluation of recursive rules which is exploited in translating the until
and release formulas. 

For simple temporal properties such as reachability and
deadlock detection our approach could be quite directly
used as a basis for a similar treatment using
propositional logic and satisfiability (SAT) checkers.
This is fairly straightforward by using the ideas of Clark's
completion and  Fages' theorem~\cite{Fages94} as our encoding
produces acyclic programs except for the choice rules which need a
special treatment.

\section{Experiments}
We have implemented the deadlock detection and LTL model
checking translations presented in the previous section
in a bounded model checker
\boundsmodelsversion\ which uses
\Smodels\ as the underlying stable model finder. The implementation
performs the following optimisations
when given a fixed initial marking $M_{0}$:
\begin{itemize}
\item Place and transition atoms are added only
from the time step they can first appear on. Only atoms for
places $p(0)$ in the initial marking are created for time $i = 0$. Then for each
$0 \leq i \leq n-1$: (i) Add transition atoms for all transitions $t(i)$
such that all the place atoms in the preset of $t(i)$ exist. (ii)~Add
place atoms for all places $p(i+1)$ such that either the place atom
$p(i)$ exists or some transition atom in the preset of $p(i+1)$ exists.
\item Duplicate rules are removed. Duplicates can appear in
(\ref{eq:conflicts}) and (\ref{eq:live}).
\end{itemize}

We compare \boundsmodels\ to a state of the art model checker
\NuSMVversion\ (\url{http://nusmv.irst.itc.it/})
which contains two different model checking engines~\cite{NuSMV:cav02}.
The first one (\NuSMVBMC) is a bounded LTL model checker based on
the approach of~\cite{BiereCimattiClarkeZhu:TACAS1999},
and includes some further improvements presented in~\cite{NuSMV:BMC}.
It uses as the underlying SAT solver the \zChaffversion\
(\url{http://www.ee.princeton.edu/~chaff/})
system \cite{Chaff}.
The second engine (\NuSMVBDD) is an efficient implementation of
a traditional BDD based model checker.

As benchmarks we use a set of deadlock detection benchmarks collected
by Corbett~\cite{Corbett94:deadlock}, and also hand-crafted LTL model
checking problems based on these models. The Corbett models are
available both as communicating automata, and in the input language of the
\NuSMV\ model checker.
The communicating automata models were converted
to 1-safe P/T-nets by Melzer and R{\"o}mer~\cite{MR:cav97}.
We use the models which have a deadlock, and are non-trivial to model check.

In deadlock checking experiments
for each model and both semantics we increment the
used bound until a deadlock is found. We report
the time for \Smodelsversion\ to find the first stable model using this bound
and the time used by the
\NuSMV\ model checker.
In some cases a model could not be found within a reasonable time
(3600 seconds) in which case we report the time used
to prove that there is no deadlock within the reported bound.

\begin{table}
\begin{center}
\caption{Deadlock Checking Experiments}\label{table:experiments1}
{\scriptsize
\begin{tabular}{l@{}rrrrrrrr}\hline\hline
Problem & St $n$ & St $s$ & Int $n$ & Int $s$ & Bmc $n$ & Bmc $s$ & Bdd $s$ & States \\
\hline
{DP(6)}& 1 & 0.0 & 6 & 0.1 & 6 & 0.2 & 0.1 & 728\\
{DP(8)}& 1 & 0.0 & 8 & 2.3 & 8 & 2.4 & 0.1 & \corr{6560}\\
{DP(10)}& 1 & 0.0 & 10 & 182.5 & 10 & 155.9 & 0.2 & \corr{59048}\\
{DP(12)}& 1 & 0.0 & $>$9 & 707.3 & $>$8 & 984.4 & 0.2 & \corr{{531440}}\\
\hline
{KEY(2)}& $>$29 & 2089.7 & $>$29 & 2227.8 & $>$30 & 2531.9 & 0.1 & 536\\
\hline
{MMGT(3)}& 7 & 0.9 & 10 & 24.2 & 10 & 16.6 & 0.2 & 7702\\
{MMGT(4)}& 8 & 174.9 & 12 & 2533.4 & 12 & 84.9 & 0.4 & 66308\\
\hline
{Q(1)}& 9 & 0.0 & $>$17 & 1051.4 & $>$11 & 2669.8 & 2.9 & 123596\\
\hline
{DARTES(1)\,}& 32 & 0.4 & 32 & 0.4 &  $*$ & $*$ & $*$ & \corr{{$>$1500000}}\\
\hline
{ELEV(1)}& 4 & 0.0 & 9 & 0.1 & - & - & - & \corr{163}\\
{ELEV(2)}& 6 & 0.2 & 12 & 1.8 & - & - & - & \corr{1092}\\
{ELEV(3)}& 8 & 1.9 & 15 & 94.2 & - & - & - & \corr{7276}\\
{ELEV(4)}& 10 & 60.9 & $>$13 & 656.8 & - & - & - & \corr{48217}\\
\hline
{HART(25)}& 1 & 0.0 & $>$5 & 0.4 & - & - & - & \corr{{$>$1000000}}\\
{HART(50)}& 1 & 0.0 & $>$5 & 1.7 & - & - & - & \corr{{$>$1000000}}\\
{HART(75)}& 1 & 0.0 & $>$5 & 5.1 & - & - & - & \corr{{$>$1000000}}\\
{HART(100)}& 1 & 0.0 & $>$5 & 11.6 & - & - & - & \corr{{$>$1000000}}\\
\hline\hline
\end{tabular}
}
\end{center}
\end{table}

The deadlock checking experimental results can be found in Table~\ref{table:experiments1}.
We use ``$*$'' to denote the fact that
\NuSMV\ ran out of 900MiB memory limit on DARTES(1)
with both engines, so we could not make a comparison in this case.
While performing state space size comparisons between Petri net and
\NuSMV\ models, we found problems in the used communicating automata to
Petri net translation, resulting in model differences in ELEV(x) and HART(x).
Thus we also excluded these models from comparison denoting this
in the table with ``-''.

The columns are:
\begin{itemize}
\item Problem: The problem name with the size of the instance
in parenthesis.
\item St~$n$: The smallest integer $n$ such that a deadlock could be found
using the step semantics / in case of $>n$ the largest integer
$n$ for which we could prove that there is no deadlock within that bound
using the step semantics.
\item St~$s$: The time in seconds to find the first stable model / to prove
that there is no stable model. (See St~$n$ above.) 
\item Int~$n$ and Int~$s$: defined as St~$n$ and St~$s$ but for the
interleaving semantics. 
\item Bmc~$n$ and Bmc~$s$: Same as Int~$n$ and Int~$s$ above, but for
the \NuSMVBMC\ bounded model checking engine.
\item Bdd~$s$: Time needed for the
\NuSMVBDD\ engine to compute the set of reachable
states and to find a state in that set which has no successors.
\item States: Number of reachable states of the model
(if known).
\end{itemize}
The time reported is the average of 5 runs where the timing is
measured by the {\tt /usr/bin/time} command
on a 1GiB RAM, 1GHz AMD (Thunderbird) Athlon PC running Linux.
The time needed for creating the \Smodels\ input was very small,
and therefore omitted.

The \NuSMVBMC\ engine
did not directly support deadlock checking, so
we had to modify the models slightly to add a proposition $\mathit{Live}$
to all the models, which is true iff any transition is enabled.
We then ask for counterexamples without a loop for the LTL property
$\Box \mathit{Live}$. With the \NuSMVBDD\ engine we use forward reachability
checking combined with transition relation totality check limited to
the reachable states.
The default dynamic variable reordering method is used.
We disable for these deadlock checking experiments a time consuming
(and unnecessary for deadlock checking) fairness set calculation
during
\NuSMVBDD\ model initialisation.

When comparing our bounded model checker on step and interleaving semantics
we note that in many of the experiments the step semantics version finds
a deadlock with a smaller bound than the interleaving one.
Also, when the bound needed to find the deadlock is fairly
small, our bounded model checker is performing well.
In the examples ELEV(4), HART(x) and Q(1) we are able to find a 
counterexample only when using step semantics. 
In the KEY(2) example we are not able to find a counterexample
with either semantics, even though the problem is known to
have only a small number of reachable states. In contrast, the
DARTES(1) problem has a large state-space, and despite of it
a counterexample of length 32 is obtained. 

When comparing with
\NuSMVBMC\ we observe that the step semantics
translation is quite competitive, with only
\NuSMVBMC\ being better on KEY(2)
and MMGT(4). We believe this is mainly due to the smaller bounds obtained
using steps. Somewhat surprisingly to us,
\NuSMVBMC\ is also worse than
interleaving on DP(12) and Q(1).
This could be due to either translation or solver differences.

The examples we have used have a small and fairly regular state space.
Thus the \NuSMVBDD\ engine
is very competitive on them, as expected.
The only exception to this rule is DARTES(1), where for some reason
the
\NuSMVBDD\ engine uses more than 900 MiB of memory.
Overall, the results are promising, in
particular, for small bounds and the step semantics.

\begin{table}
\begin{center}
\caption{LTL Model Checking Experiments}\label{table:experiments2}
{\scriptsize
\begin{tabular}{l@{}rrrrrrrr}\hline\hline
Problem & St $n$ & St $s$ & Int $n$ & Int $s$ & Bmc $n$ & Bmc $s$ & Bdd $s$ & States \\
\hline
{DP(6)}& 7 & 0.2 & 8 & 0.5 & 8 & 4.3 & 64.8 & 728\\
{DP(8)}& 8 & 1.5 & 10 & 5.7 & 10 & 64.0 & $>$1800 & \corr{6560}\\
{DP(10)}& 9 & 25.9 & 12 & 140.1 & 12 & 1257.1 & $>$1800 & \corr{59048}\\
{DP(12)}& 10 & 889.4 & 14 & $>$1800 & 14 & $>$1800 & $>$1800 & \corr{{531440}}\\
\hline\hline
\end{tabular}
}
\end{center}
\end{table}

We do not have a large collection of LTL model checking problems
available to us. Instead we pick a model family, the dining
philosophers problems DP(x), and use a hand-crafted  LTL formula for each
model. Because the
\NuSMVBDD\ LTL model checking engine only works
for deadlock free models, we remove all the deadlocks from these examples
by making each deadlock state a~successor of itself.

The formulas to be checked are hand-crafted to demonstrate
potential differences
between~\cite{BiereCimattiClarkeZhu:TACAS1999} and
our proposed method.
We study nested until formulas for which the translation
of~\cite{BiereCimattiClarkeZhu:TACAS1999} 
seems to be rather complex.
In~our model
the atomic proposition $f_i.\mathit{up}$ has the meaning that fork $i$
is available, and $p_i.\mathit{eat}$ has the meaning that philosopher
$i$ is eating. We model check the following formulas.
For six philosophers we use the formula:
$$\neg \Box \Diamond (f_5.\mathit{up} \; U \; (p_5.\mathit{eat} \wedge (f_3.\mathit{up} \; U \; (p_3.\mathit{eat} \wedge (f_1.\mathit{up} \; U \; p_1.\mathit{eat}))))),$$
for eight philosophers we use the formula:
$$\neg \Box \Diamond (f_7.\mathit{up} \; U \; (p_7.\mathit{eat} \wedge (f_5.\mathit{up} \; U \; (p_5.\mathit{eat} \wedge (f_3.\mathit{up} \; U \; (p_3.\mathit{eat} \wedge (f_1.\mathit{up} \; U \; p_1.\mathit{eat}))))))),$$
and so on.
The counterexample is a model for a formula of the form $\Box
\Diamond (\varphi)$, 
where $\varphi$ has deeply nested until formulas.
Thus in a counterexample $\varphi$ has to hold infinitely often.
As an example,
one way to make $\varphi$ hold in the six philosophers case is to find a state where
($p_5.\mathit{eat} \wedge p_3.\mathit{eat} \wedge p_1.\mathit{eat}$)
holds.

The experimental results for the LTL model checking
can be found in Table~\ref{table:experiments2}. For this set of experiments
we use the run time limit of 1800 seconds, and do not try smaller bounds
when the limit is exceeded. The columns of the table are as in deadlock
checking experiments, except that we are looking for a counterexample
to the LTL formula.
In these examples \NuSMV\ is run
with default dynamic BDD variable reordering on.

The experiments show that the step semantics is able to obtain a counterexample
for DP(12), while other methods are unable to. The
\NuSMVBMC\ engine scales
worse than the interleaving semantics translation. By investigating further, we
notice that in DP(10)
the \zChaff\ solver only takes 160.4 seconds, while
the generation of the SAT instance for the solver takes almost 1100 seconds.
We believe that a large part of this overhead is due to the size of the
generated LTL model checking translation.
The \NuSMVBDD\ based LTL model checker
seems to be scaling worse than for the corresponding deadlock checking
examples and 
it can be observed that the number of BDD operations required for LTL
model checking is significantly larger.

The used tools, models, formulas, and logic programs are available at\\
\url{http://www.tcs.hut.fi/~kepa/experiments/boundsmodels/}.

\section{Conclusions}

We introduce bounded model checking of asynchronous concurrent
systems modelled by 1-safe P/T-nets as an interesting  application area for
answer set programming. We present mappings from bounded
reachability, deadlock detection, and LTL model checking problems of
1-safe P/T-nets to stable model computation.
Our approach is capable of doing model checking for a set of initial
markings at once. This is usually difficult to achieve in current
enumerative model checkers and often leads to state space explosion.
We handle asynchronous systems using a step semantics whereas previous
work on bounded model checking only uses the interleaving
semantics~\cite{BiereCimattiClarkeZhu:TACAS1999}. Furthermore, our
encoding is more compact than the previous approach
employing propositional
satisfiability~\cite{BiereCimattiClarkeZhu:TACAS1999}. 
This is because our rule based approach allows to represent executions
of the system, e.g.\ frame axioms, succinctly and supports directly the
recursive fixed point computation needed to evaluate LTL formulas. 
Another feature of our LTL translation is that
it does not require the deadlock freeness
assumption used by~\cite{BiereCimattiClarkeZhu:TACAS1999},
and thus we can employ it also with systems which have not been
proved deadlock free.

The first experimental results indicate that
stable model computation is quite a competitive approach to searching for
short executions of the system leading to deadlock and worth further
study.  
More experimental work and comparisons are needed to determine the
strength of the approach. In particular, for comparing with SAT checking
techniques, it would be interesting to develop a similar treatment of asynchronous
systems using a SAT encoding and compare it to the logic program based
approach.  

Relating the net unfolding method
(see~\cite{Heljanko:Fundamenta99,MR:cav97} and further references there)
to bounded model checking would be interesting.
There are also alternative semantics to the
two presented in this work~\cite{Heljanko:Concur2001},
applying them to bounded LTL model checking is left for further work.

\bibliographystyle{alpha}
\bibliography{ltl-bmc}

\appendix
\section{Proofs}

\subsection{Proof of Theorem \ref{th:reachability}\label{app:reachability}}

We first recall our proof objective.
Let $N = \seq{P, T, F}$ be a 1-safe P/T-net for all initial markings
satisfying a condition~$C_0$.

We want to prove that the
net $N$ has an initial marking satisfying $C_0$
such that a marking satisfying a condition $C$ is reachable in at most $n$ steps iff
$\mprogram{C_0}{0} \cup \aprogram{N}{n} \cup \mprogram{C}{n}$ has a
stable model.

The proof
is based on the following
two lemmata which establish a correspondence between stable models of 
$\mprogram{C_0}{0} \cup \aprogram{N}{n}$ and $n$-bounded step executions of the 
1-safe P/T-net $N$. 
We say that a step
execution 
\begin{equation}
\execution{N}{n}{\smdel} =
M_0~\overset{S_{0}}{\rightarrow}~M_1~\overset{S_{1}}{\rightarrow}~\ldots
M_{n-1}~\overset{S_{n-1}}{\rightarrow}~M_{n}
\label{eq:execution}
\end{equation}
 is 
\emph{derived from} a stable model $\smdel$
if for all $i = 0,\ldots,n$, $M_i = \{p \in P \mid p(i) \in \smdel\}$
and for all $i = 0,\ldots,n-1$, $S_i = \{t \in T \mid t(i) \in
\smdel\}$. 
\begin{lemma}
  Let $N = \seq{P, T, F}$ be a 1-safe P/T-net for all initial markings
  satisfying a condition $C_0$.
If $\mprogram{C_0}{0} \cup \aprogram{N}{n}$ has a stable model
$\smdel$, then $\execution{N}{n}{\smdel}$ is a step execution of $N$ 
starting from an initial marking satisfying $C_0$. 
\label{le:model:to:execution}
\end{lemma}

\begin{proof}
Consider a step execution $\execution{N}{n}{\smdel}$ (\ref{eq:execution})
which is derived from a stable model $\smdel$ of $\mprogram{C_0}{0} \cup
\aprogram{N}{n}$.  
Because $\smdel$ satisfies rules $\mprogram{C_0}{0}$, then marking $M_0$
satisfies condition $C_0$. 
Now we show that
$\execution{N}{n}{\smdel}$ is a valid step execution starting from
$M_0$ by showing that if the step execution is valid up to marking $M_i$,
then it is valid also up to $M_{i+1}$, i.e.,
$M_i~\overset{S_{i}}{\rightarrow}~M_{i+1}$ holds for all $i = 0,\ldots,n-1$.
Consider $S_i = \{t \in T \mid t(i) \in \smdel\}$. 
As every stable model is supported, $t(i) \in \smdel$ implies that there is
a rule in $\mprogram{C_0}{0} \cup \aprogram{N}{n}$ with $t(i)$ as the
head and the
body literals satisfied in $\smdel$. The only candidate rule is
(\ref{eq:tchoices}) and, hence, 
for every place $p$ in the preset of $t$, 
$p(i) \in \smdel$ and, thus, $p \in M_i$.
This implies that every transition
$t \in S_i$ is enabled in $M_i$. Moreover, as $\smdel$ satisfies
rules~(\ref{eq:conflicts}), $S_i$ is concurrently enabled in $M_i$. 

Given a marking $M_i$ and a concurrently enabled step $S_i$, 
$M_i~\overset{S_{i}}{\rightarrow}~M_{i+1}$ holds in a 1-safe net,
if for all $p \in P$, $ p \in M_{i+1}$ iff 
\begin{equation}
\mbox{(a) } p \in t^\bullet \mbox{ for some } t \in
S_i \mbox{ or } 
\mbox{(b) } 
p \in M_i \mbox{ and for all } t \in S_i, p \not\in {}^\bullet t .
\label{eq:step}
\end{equation}
We complete the proof by showing that this holds for $M_{i+1}$. 
Consider a place $p \in P$. 

($\Rightarrow$) If $p \in M_{i+1}$, then $p(i+1) \in \smdel$. Hence,
there is some rule in $\aprogram{N}{n}$ with $p(i+1)$ as the
head and the body literals satisfied in $\smdel$. 
There are two types of candidate rules (\ref{eq:effects}) and (\ref{eq:frame}). 
In the case of (\ref{eq:effects}), if the body is satisfied in $\smdel$,
$t(i) \in \smdel$ and $t \in S_i$ for a transition $t$ with 
$p \in t^\bullet$ implying that Condition (\ref{eq:step}:~a) holds. 
For (\ref{eq:frame}), if the body is satisfied in $\smdel$,
then $p \in M_i$ and no transition having $p$ in its preset is in
$S_i$. This implies that Condition (\ref{eq:step}:~b) holds.

($\Leftarrow$) If Condition (\ref{eq:step}:~a) holds for $p \in P$, then
there is some $t(i) \in \smdel$. Because a rule 
$p(i+1) \lparrow t(i)$ of type (\ref{eq:effects}) is in
$\aprogram{N}{n}$, 
$p(i+1)\in \smdel$ and, hence, $p \in
M_{i+1}$.  If Condition (\ref{eq:step}:~b) holds for $p \in P$, then
$p(i) \in \smdel$ and for all transition $t$ with $p \not\in {}^\bullet
t $, $t(i) \not\in \smdel$. As $\smdel$ satisfies a rule 
(\ref{eq:frame}) for $p(i+1)$, $p(i+1)\in \smdel$ and, hence, $p \in
M_{i+1}$.
\end{proof}

\begin{lemma}
  Let $N = \seq{P, T, F}$ be a 1-safe P/T-net for all initial markings
  satisfying a condition $C_0$.
  If there is a step execution $\sigma'$ of $N$ without empty steps
  from an initial marking $M_0$ satisfying
  $C_0$ containing $n' \leq n$ steps, then
  there is a stable model $\smdel$ of 
  $\mprogram{C_0}{0} \cup \aprogram{N}{n}$ such that the derived step
  execution $\sigma = \execution{N}{n}{\smdel} = 
  M_0~\overset{S_{0}}{\rightarrow}~M_1~\overset{S_{1}}{\rightarrow}~\ldots
  M_{n-1}~\overset{S_{n-1}}{\rightarrow}~M_{n}$ is a step execution of $N$,
  such that $\sigma$ is the execution $\sigma'$ with $n -n'$ empty steps
  added to the beginning.
\label{le:execution:to:model}
\end{lemma}
\begin{proof} 
Let $\sigma'$ be a step execution from an initial marking $M_0$ satisfying
$C_0$ in $n'\leq n$ steps.
Then there is  a step execution 
\begin{equation}
M_0~\overset{S_{0}}{\rightarrow}~M_1~\overset{S_{1}}{\rightarrow}~\ldots
M_{n-1}~\overset{S_{n-1}}{\rightarrow}~M_{n}
\end{equation}
where $n-n'$ first steps are empty if $n'<n$, i.e., $S_0=\cdots=S_{n-n'-1} = \{\}$
and $M_0=\cdots=M_{n-n'}$. 

Now consider a set of atoms 
\begin{eqnarray*}
\smdel & = & \{ p(i) \mid p \in M_i, 0\leq i \leq n \} \cup \\
         & & \{ t(i) \mid t \in S_i, 0\leq i < n \} \cup \\
         && \{\mathit{idle}(0),\ldots,\mathit{idle}(n-n'-1)\} \cup \\
&& \{p'(0) \mid p \in P -M_0\} \cup \{ t'(i) \mid t \in T- S_i, 0\leq i
         < n \} \cup C(0) 
\end{eqnarray*}
where $C(0)$ are the atoms $c_f(0)$ corresponding to the subexpressions
$f$ of Condition $C_0$ that are satisfied in $M_0$. 

We show that $\smdel$ is a stable model of $\myprogram
=\mprogram{C_0}{0} \cup \aprogram{N}{n}$ by establishing that 
(i) $\smdel \models \reduct{\myprogram}{\smdel}$ and that 
(ii) if $\smdel' \subseteq \smdel$ and $\smdel' \models
\reduct{\myprogram}{\smdel}$, then $\smdel \subseteq \smdel'$ 
which together imply that $\smdel$ is the
minimal set of atoms satisfying $\reduct{\myprogram}{\smdel}$.

(i) By construction the rules in $\reduct{\mprogram{C_0}{0}}{\smdel}$ are
satisfied by $\smdel$. Now we consider other rules in $\aprogram{N}{n}$ 
case by case and show that rules resulting from them in
$\reduct{\myprogram}{\smdel}$ are satisfied. 
Rules resulting from (\ref{eq:initialmarking}) and (\ref{eq:tchoices}) 
are satisfied directly by construction of $\smdel$ because 
$p(0) \not\in\smdel$ iff $p'(0) \in\smdel$ and 
$t(i) \not\in\smdel$ iff $t'(i) \in\smdel$. 
Consider a rule (\ref{eq:effects}) and assume that $t(i) \in
\smdel$. Now $t \in S_i$ with $p \in t^\bullet$. This implies that $p
\in M_{i+1}$ and, hence, $p(i+1) \in \smdel$. 
Each rule (\ref{eq:conflicts}) is satisfied by $\smdel$ because each $S_i$
is concurrently enabled implying that no $S_i$ can contain any two
transition sharing place in their presets. 
Consider the reduct $p(i+1) \lparrow p(i) \in
\reduct{\myprogram}{\smdel}$  of a rule (\ref{eq:frame}) and 
the case where $p(i) \in \smdel$. Now  $p \in M_i$ and 
for each transition $t$ with $p$ in its postset $t(i) \not\in \smdel$. Hence,
there is no transition with $p$ in its
preset in $S_i$ implying that $p \in M_{i+1}$ and $p(i+1) \in \smdel$. 
Rules (\ref{rule:idlestart}) are straightforwardly satisfied by
construction of $\smdel$. 
Hence, $\smdel \models \reduct{\myprogram}{\smdel}$ holds. 

(ii) Consider a set $\smdel' \subseteq \smdel$ such that
$\smdel' \models \reduct{\myprogram}{\smdel}$. Assume that there is 
an atom $x \in \smdel-\smdel'$. This atom cannot be 
any $p(0)$ for a place $p$ because for each $p(0) \in \smdel$ there is
a fact $p(0) \lparrow \in \reduct{\myprogram}{\smdel}$.
Similarly, it cannot be any $p'(0)$, $t'(i)$ for some $t \in T$ or
$\mathit{idle}(i)$ because also for each of these there is a
corresponding fact in $\reduct{\myprogram}{\smdel}$. 

Hence, $x$ is either 
some $p(i)$ with $p \in P$ and $0 < i \leq n$ or 
some $t(i)$ with $t \in T$ and $0 \leq i \leq n$. 
Now consider such an atom $x$ with the smallest index $i$.  Suppose $x$
is some $p(i) \in \smdel-\smdel'$.  Then $p \in M_i$ which implies that 
(a) $p \in t^\bullet \mbox{ for some } t \in S_{i-1}$ or
(b) $p \in M_{i-1} \mbox{ and for all } t \in S_{i-1}, p \not\in {}^\bullet t$.
In the case (a) there is some $t(i-1) \in  \smdel'$ and as 
$\smdel'$ satisfies a rule of type (\ref{eq:effects}) for $p(i)$, $p(i)
\in\smdel'$. 
In the case (b), $p(i) \lparrow p(i-1) \in
\reduct{\myprogram}{\smdel}$ and $p(i-1) \in \smdel'$ which implies
$p(i) \in\smdel'$. 
Hence, in both cases $p(i) \in\smdel'$ holds implying that $x$ must be
some $t(i)$ with $t \in T$ and $0 \leq i \leq n$.  
As $t(i) \in \smdel$, $t \in S_i$ implying that $t$ is enabled and, 
hence, that every place $p$ in the preset of $t$ is in $M_{i}$. But
then  for every place $p$ in the preset of $t$, $p(i) \in \smdel$ and,
hence, $p(i) \in \smdel'$. As $\smdel'$ satisfies the rule 
$t(i) \lparrow p_1(i), \ldots, p_l(i) \in \reduct{\myprogram}{\smdel}$
where $\{p_1, \ldots, p_l\}$ is 
the preset of $t$, $t(i) \in \smdel'$, a contradiction. 
Thus, $\smdel \subseteq \smdel'$.
\end{proof}

\begin{proof} (of Theorem~\ref{th:reachability}).

($\Leftarrow$) 
If $\mprogram{C_0}{0} \cup \aprogram{N}{n} \cup \mprogram{C}{n}$ has a
stable model $\smdel$, then by Proposition~\ref{pro:projection}
there is a stable model $\smdelE = \smdel \cap At$ of 
$\mprogram{C_0}{0} \cup \aprogram{N}{n}$ 
where 
$At = \atoms{\mprogram{C_0}{0} \cup \aprogram{N}{n}}$. 
By Lemma~\ref{le:model:to:execution}
$\execution{N}{n}{\smdelE}$ is a step execution of $N$ 
starting from an initial marking satisfying $C_0$.
As $\smdel$ satisfies rule $\mprogram{C}{n}$, then 
the marking $M_n$ in $\execution{N}{n}{\smdelE}$ satisfies 
condition $C$. 

($\Rightarrow$) 
If $N$ has an initial marking $M_0$ satisfying $C_0$
such that a marking $M$ satisfying condition $C$ is reachable in at most
$n$ steps, then by Lemma~\ref{le:execution:to:model}
$\mprogram{C_0}{0} \cup \aprogram{N}{n}$ has a stable model 
$\smdelE$ such that the derived step
  execution $\execution{N}{n}{\smdelE} = 
  M_0~\overset{S_{0}}{\rightarrow}~M_1~\overset{S_{1}}{\rightarrow}~\ldots
  M_{n-1}~\overset{S_{n-1}}{\rightarrow}~M_{n}$ is a step execution of $N$
  and $M=M_n$.
The rules $\mprogram{C}{n} - \{\lparrow \lpnot c(n)\}$ are stratified
and the heads do not occur in $\mprogram{C_0}{0} \cup \aprogram{N}{n}$. 
By Proposition~\ref{pro:strat:extension} there is a unique stable model
$\smdel$ 
of $\mprogram{C_0}{0} \cup \aprogram{N}{n} \cup \mprogram{C}{n} -
\{\lparrow \lpnot c(n) \}$ such that $\smdelE = \smdel \cap At$ where
$At = \atoms{\mprogram{C_0}{0} \cup \aprogram{N}{n}}$.
As $M_n$ satisfies condition $C$, then $c(n) \in \smdel$ and, hence, 
$\lparrow \lpnot c(n)$ is satisfied by $\smdel$ implying by
Proposition~\ref{pro:ic:extension} that $\smdel$ is a stable model of 
$\mprogram{C_0}{0} \cup \aprogram{N}{n} \cup \mprogram{C}{n}$.
This concludes our proof of Theorem~\ref{th:reachability}.\end{proof}

\subsection{Proof of Theorem \ref{th:ltl}\label{app:ltl1}}

We first recall our proof objective.
Let $f$ be an LTL formula in positive normal form and
$N = \seq{P, T, F}$ be a 1-safe P/T-net for all initial markings
satisfying a condition $C_0$.

We want to prove that whenever we have a stable model
$\smodel{LTL}$ of the program
$\myprogram = \mprogram{C_0}{0} \cup \aprogram{N}{n} \cup \lprogram{N}{n} \cup \ltlprogram{f}{n}$
we can construct a maximal execution of the
net system $N$ from an initial marking satisfying $C_0$ which satisfies~$f$.

Our proof proceeds as follows.
We first derive a step execution $\sigma'$ from the stable model $\smodel{LTL}$.
We then create a maximal step execution $\sigma''$ from $\sigma'$ using
an index $0 \leq l \leq n$ also obtained from $\smodel{LTL}$.
After this we show that a maximal (interleaving) execution $\sigma$ can
be obtained from $\sigma''$ such that $\sigma \models f$ iff $\sigma'' \models f$.
Finally we show that $\sigma \models f$.

\begin{lemma}
For the stable model $\smodel{LTL}$, there is
a step execution $\sigma'$ of the net system $N$ from
an initial marking satisfying $C_0$.
\label{le:stepex}
\end{lemma}
\begin{proof}
We first use Proposition~\ref{pro:projection} with $\myprogram_1 = \mprogram{C_0}{0} \cup \aprogram{N}{n}$
and $\myprogram_2 = \lprogram{N}{n} \cup \ltlprogram{f}{n}$
to obtain a stable model $\smodel{1}$ of the subprogram $\myprogram_1$.
By Lemma~\ref{le:model:to:execution} the execution
$\sigma' = \execution{N}{n}{\smodel{1}}$ is a step execution of $N$ 
starting from an initial marking satisfying $C_0$.
\end{proof}

Let $\smodel{1}$ be the stable model and $\sigma'$ the step execution
obtained in the proof of Lemma~\ref{le:stepex} above. Now we show
that by adding the rules $\lprogram{N}{n}$ we can evaluate
whether the last marking reached by $\sigma'$ is a deadlock.

\begin{lemma}
Let  $\smodel{1}$ be a stable model of $\myprogram_1 = \mprogram{C_0}{0}
\cup \aprogram{N}{n}$ and 
$\smodel{2}$ a stable model of $\myprogram_2 = \myprogram_1
 \cup \lprogram{N}{n}$.
Then $\mathit{live} \in \smodel{2}$ iff the last marking
reached by $\execution{N}{n}{\smodel{1}}$ is {\em not} a deadlock.
\label{le:livelockok}
\end{lemma}

\begin{proof}
The rules in $\lprogram{N}{n}$ are stratified. Hence, 
by Proposition~\ref{pro:strat:extension} $\smodel{2}$
is the unique stable model of $\myprogram_2$ such that $\smodel{1} =
\smodel{2} \cap \atoms{\myprogram_1}$. 
If $\mathit{live} \in \smodel{2}$, then there is some rule in
$\lprogram{N}{n}$ with its body satisfied by $\smodel{2}$. As 
$\smodel{1} = \smodel{2} \cap \atoms{\myprogram_1}$, the body is
satisfied by $\smodel{1}$ and, hence, there is an enabled transition in
the last marking reached by $\execution{N}{n}{\smodel{1}}$. 
In the other direction, if there is an enabled transition $t$ in
the last marking, then $\{ p_1(n),\ldots,p_l(n) \} \subseteq
\smodel{1}$ where $\{ p_1,\ldots,p_l \}$ is the preset of $t$. 
But then $\{ p_1(n),\ldots,p_l(n) \} \subseteq \smodel{2}$ and
$\mathit{live} \in \smodel{2}$. 
\end{proof}

Hence, we can again use Proposition~\ref{pro:projection} with
$\myprogram_1 = \mprogram{C_0}{0} \cup \aprogram{N}{n} \cup \lprogram{N}{n}$ and
$\myprogram_2 = \ltlprogram{f}{n}$ together with Lemma~\ref{le:livelockok}
to show that $\mathit{live} \in \smodel{LTL}$ iff the last marking
reached by $\sigma'$ is {\em not} a deadlock.

We now do a case analysis on three different types of counterexamples.
The stable model $\smodel{LTL}$ belongs to exactly one
of the following three mutually exclusive cases:
\begin{itemize}
\item[a)] $\mathit{nl}(l+1) \in \smodel{LTL}$ for some $0 \leq l \leq
  n-1$: infinite maximal execution which we will represent as a
pair $(\sigma', l)$, where $0 \leq l \leq n-1$ such that
$\mathit{nl}(l+1) \in \smodel{LTL}$,
\item[b)] $\mathit{nl}(i+1) \not \in \smodel{LTL}$ for all $0 \leq i \leq n-1$,
$\mathit{live} \not \in \smodel{LTL}$: finite maximal execution which 
we will represent  as a pair $(\sigma', n)$, or
\item[c)] $\mathit{nl}(i+1) \not \in \smodel{LTL}$ for all $0 \leq i \leq n-1$,
$\mathit{live} \in \smodel{LTL}$: non-maximal execution which
we will also represent as a pair $(\sigma', n)$.
\end{itemize}

We will now analyse the stable model $\smodel{LTL}$.
\begin{lemma}
The following holds for the three different cases of $\smodel{LTL}$.
\begin{itemize}
\item[a)] If $\mathit{nl}(l+1) \in \smodel{LTL}$, then 
$\mathit{nl}(i) \not \in \smodel{LTL}$
for all $i \not = (l + 1)$,
$\mathit{le} \in \smodel{LTL}$,
$\mathit{il}(i) \in \smodel{LTL}$ for all $l+1 \leq i \leq n$ and
$\mathit{live} \in \smodel{LTL}$.
\item[b)] If $\mathit{nl}(i+1) \not \in \smodel{LTL}$ for all $0 \leq i
  \leq n-1$ and $\mathit{live} \not \in \smodel{LTL}$, then 
$\mathit{le} \not \in \smodel{LTL}$, and
$\mathit{live} \not \in \smodel{LTL}$.
\item[c)] If $\mathit{nl}(i+1) \not \in \smodel{LTL}$ for all $0 \leq i
  \leq n-1$, and $\mathit{live} \in \smodel{LTL}$, 
  then $\mathit{le} \not \in
  \smodel{LTL}$. 
\end{itemize}
\label{le:compatible}
\end{lemma}
\begin{proof}
\begin{itemize}
\item[a)] Because the only rule
with $\mathit{nl}(l+1)$ as head is
$\mathit{nl}(l+1) \lparrow {el}(l)$ we also get that
$\mathit{el}(l) \in \smodel{LTL}$.
Because rule (\ref{rule:guesscheck}) is satisfied
we know that for all $i \not = l$ it holds that $\mathit{el}(i) \not \in \smodel{LTL}$.
Because $\mathit{le} \lparrow \mathit{el}(l) \in \myprogram$,
we also know that $\mathit{le} \in \smodel{LTL}$.
By using the rules
$\mathit{il}(i+1) \lparrow \mathit{el}(i)$ and $\mathit{il}(i+1) \lparrow \mathit{il}(i)$
and the fact that $\mathit{el}(l) \in \smodel{LTL}$
combined with simple induction we get that
that $\mathit{il}(i) \in \smodel{LTL}$ for all $l + 1 \leq i \leq n$.
The rules (\ref{rule:checkloop}) imply that 
$p \in M_{l}$ iff $p \in M_{n}$.
From the rule (\ref{rule:noidleloop}) and $\mathit{le} \in \smodel{LTL}$
we get that
$\mathit{idle}(n -1) \not \in \smodel{LTL}$
and thus the step $S_{n-1}$ is non-empty.
Taking together that $M_{l} = M_{n}$ and that the step
$S_{n-1}$ is non-empty
implies $M_{n}$ is not a deadlock, and thus
$\mathit{live} \in \smodel{LTL}$.

\item[b,c)] Because $\mathit{nl}(i+1) \not \in \smodel{LTL}$ for all $0 \leq i \leq n-1$,
we know also that $\mathit{el}(i) \not \in \smodel{LTL}$ for all $0 \leq i \leq n-1$.
Then as the only rules having $\mathit{le}$ as head are the rules (\ref{rule:loopstuff})
of the form $\mathit{le} \lparrow \mathit{el}(i)$, 
$\mathit{le} \not \in \smodel{LTL}$.
\end{itemize}
\end{proof}

We record some facts discovered in the proof of
Lemma~\ref{le:compatible}, case a), in the following.
\begin{corollary}
In the case a) for $\sigma'$ it holds that 
$M_{l} = M_n$  and the step $S_{n-1}$ is non-empty.
\label{co:loop}
\end{corollary}

We will next state an additional property of $\sigma'$ and 
show how a maximal step execution $\sigma''$ can be obtained
given the pair $(\sigma', l)$.

\begin{lemma}
Each step of $\sigma'$ contains at most one visible transition.
\label{le:onevis}
\end{lemma}
\begin{proof}
We use Proposition~\ref{pro:ic:extension}
with the rules~(\ref{rule:onevisible}) of the subprogram
$\ltlprogram{f}{n}$.
\end{proof}

\begin{lemma}
For the stable model $\smodel{LTL}$ there is 
a maximal step execution $\sigma''$ of the net system $N$ from
an initial marking satisfying $C_0$.
\label{le:maxstepex}
\end{lemma}

\begin{proof}
In all cases below $M_0$ is the initial marking of $\sigma'$ and
thus satisfies $C_0$.

In the case a)
we know from Corollary~\ref{co:loop} that $M_{l} = M_n$
and we can thus generate an infinite maximal step execution of
$N$ using the pair $(\sigma',l)$. The corresponding
infinite step execution $\sigma''$ is
\[
M_0~\overset{S_{0}}{\rightarrow}~M_1~\overset{S_{1}}{\rightarrow}~\cdots
M_{n-1}~\overset{S_{n-1}}{\rightarrow}~M_{n}~\overset{S_{l}}{\rightarrow}~M_{l+1}~\overset{S_{l+1}}{\rightarrow}~\cdots
M_{n-1}~\overset{S_{n-1}}{\rightarrow}~M_{n}~\overset{S_{l}}{\rightarrow}
\cdots 
\]
We also know from Corollary~\ref{co:loop} that
the step $S_{n-1}$ is non-empty. Therefore $\sigma''$
contains infinitely many non-empty steps.

In the case b) the step execution $\sigma'' = \sigma'$ will be a maximal step
execution of N.

In the case c) we can pick an {\em interleaving} execution $\sigma'''$
such that the concatenation of $\sigma'$ followed by $\sigma'''$
will be a maximal step execution $\sigma''$ of $N$.
\end{proof}

We can now state the existence of maximal executions given the stable
model $\smodel{LTL}$.
\begin{lemma}
For the stable model $\smodel{LTL}$ there is 
a maximal (interleaving) execution $\sigma$ of the net system $N$ from
an initial marking satisfying $C_0$.
\label{le:maxex}
\end{lemma}

\begin{proof}
By using the procedure described above we can obtain the maximal step execution
$\sigma''$ from $\smodel{LTL}$.

By removing all idle time steps from the maximal step execution
$\sigma''$ of $N$, 
and replacing each step by its linearisation, i.e, by some permutation of transitions
that make up the step, we can construct a maximal {\em interleaving}
execution $\sigma$ of $N$.
The initial marking $M_0$ of $\sigma''$ is also the initial marking of $\sigma$
and thus satisfies $C_0$.
\end{proof}

Let $w, w', w'' \in V^{+} \cup V^{\omega}$ be the words corresponding
to the step executions $\sigma, \sigma', \sigma''$ discussed above,
respectively.
What we prove next is that $w \models f$ iff
$w'' \models f$ for the LTL formula $f$.

We need a technical notion of stuttering
equivalence for words. The intuition behind this equivalence is that if
two words are stuttering equivalent, they satisfy exactly the same
LTL formulas.\footnote{This property crucially
depends on the non-existence of the next-time operator $X \varphi_1$ in
our definition of LTL.}
Our definition of stuttering equivalence is motivated by a similar
definition in Chapter 10.2 of~\cite{CGP99:book} where also a longer
discussion of its use can be found.

\begin{definition}
Two words $v,v' \in V^{+} \cup V^{\omega}$ are stuttering equivalent
when:
\begin{itemize}
\item Both are infinite words $v, v' \in V^{\omega}$, and
there two infinite sequences of positive integers
$0 = i_0 < i_1 < i_2 < \ldots $
and
$0 = j_0 < j_1 < j_2 < \ldots $
such that for every $k \geq 0 :
v_{(i_{k})} = v_{(i_{k} +1)} = \ldots = v_{(i_{(k+1)} -1)} =
{v'}_{(j_{k})} = {v'}_{(j_{k} +1)} = \ldots = {v'}_{(j_{(k+1)} -1)}$, or
\item both are finite words $v,v' \in V^{+}$, and
there exist an integer $n \geq 1$ and
two finite sequences of positive integers
$0 = i_0 < i_1 < \ldots < i_{n} = |v|$
and
$0 = j_0 < j_1 < \ldots < j_{n} = |v'|$
such that for every $0 \leq k < n :
v_{(i_{k})} = v_{(i_{k} +1)} = \ldots = v_{(i_{(k+1)} -1)} =
{v'}_{(j_{k})} = {v'}_{(j_{k} +1)} = \ldots = {v'}_{(j_{(k+1)} -1)}$.
\end{itemize}
\end{definition}

The following proposition can be proved using a simple induction
on the structure of the formula $f$ using the definition of
LTL semantics.
\begin{proposition}
Let $f$ be an LTL formula and $v,v'$ be two stuttering equivalent words.
Then $v \models f$ iff $v' \models f$.
\label{pro:stutt}
\end{proposition}

\begin{lemma}
The words $w$ and $w''$ corresponding to the maximal execution $\sigma$
and the maximal step execution $\sigma''$ are stuttering equivalent.
\label{le:st:to:inter}
\end{lemma}
\begin{proof}
Lemma~\ref{le:onevis} implies that each step in $\sigma'$ consists of
at most one visible transition,
and in case c) the suffix $\sigma'''$ is interleaving by definition.
Thus each step of the execution $\sigma''$ contains at most one visible transition.
Thus when a step is replaced by some linearisation in the proof of Lemma~\ref{le:maxex},
the step is changed to
(possibly empty) stuttering of the original atomic propositions, (possibly) followed by
change in them, followed by (possibly empty) stuttering of the new atomic propositions.
Thus each step is replaced by a stuttering equivalent sequence, implying that
the whole sequence is stuttering equivalent.
\end{proof}

Lemma~\ref{le:st:to:inter} implies that if we correctly
evaluate the LTL formula $f$ for the word $w''$ then we also correctly
evaluate it for $w$. All we have to know to correctly evaluate the LTL
formula for the word $w''$ is to know the prefix
word $w'$, the index $l$, and whether we are in case a), b), or c).
The evaluation for case a) and b) will be exact, while the case c)
will only be approximate, as in that case nothing is known about
the suffix of the word $w'$.

\paragraph{Evaluating the formula $f$.}
Assume we are given a finite word
$u \in V^{+}$ of length $n+1$, index $0 \leq l \leq n$,
and knowledge whether we are in case a), b), or c).
This induces a word $y \in V^{+} \cup V^{\omega}$ such that in the case
a) $y = u (u^{(l+1)})^\omega$ and in the cases b) and c) $y = u$.

Given sufficient assumptions about a base program,
call it $\myprogram_{B}$, and its stable model $\smodel{B}$,
we want to show that by adding to it
the translation of formula $f$
as given by Fig.~\ref{fig:trans} we obtain
a program $\myprogram_{C}$, whose unique stable
model $\smodel{C}$ respects
the semantics of LTL in the following sense
for all three different cases and $0 \leq i \leq n$:
\begin{itemize}
\item[a)] $f(i) \in \smodel{C}$ iff $y^{(i)} \models f$ for $y \in V^{\omega}$ , 
\item[b)] $f(i) \in \smodel{C}$ iff $y^{(i)} \models f$ for $y \in V^{+}$, and
\item[c)] if $f(i) \in \smodel{C}$ then $y' \models f$
for all $y'' \in V^{+} \cup V^{\omega}$ such that $y' = y^{(i)} y''$.
\end{itemize}

The case c) specifies only a prefix $y$ of a word. Our encoding
cautiously under-approximates the semantics of LTL formulas
in the presence of uncertainty about the suffix $y''$.

Notice also that in the case a) the word $y$ is cyclic, and the semantics of
LTL follows the same cycle when $i \geq l$. Thus to evaluate, e.g.,
$y^{(n+1)} \models f$ it suffices to evaluate $y^{(l+1)} \models f$.

The assumptions on the program $\myprogram_B$ and its stable model $\smodel{B}$ are as follows:
\begin{enumerate}
\item The atoms appearing as heads in the LTL translation do not occur
in the program $\myprogram_B$.\label{ass:noocc}
\item For all $p \in P$, $0 \leq i \leq n$: $p(i) \in \smodel{B}$
iff $p(i) \in y_{(i)}$.\label{ass:labelok}
\item $\smodel{B}$ is exactly one of the
following three cases:
\begin{itemize}
\item[a)] $\mathit{nl}(l+1) \in \smodel{B}$, $\mathit{nl}(i) \not \in \smodel{B}$
for all $i \not = (l + 1)$,
$\mathit{le} \in \smodel{B}$,
$\mathit{il}(i) \in \smodel{B}$ for all $l+1 \leq i \leq n$, and
$\mathit{live} \in \smodel{B}$.
\item[b)] $\mathit{nl}(i+1) \not \in \smodel{B}$ for all $0 \leq i \leq n-1$,
$\mathit{le} \not \in \smodel{B}$, and
$\mathit{live} \not \in \smodel{B}$.
\item[c)] $\mathit{nl}(i+1) \not \in \smodel{B}$ for all $0 \leq i \leq n-1$,
$\mathit{le} \not \in \smodel{B}$, and
$\mathit{live} \in \smodel{B}$.
\end{itemize}
\end{enumerate}

\begin{lemma}
If the assumptions stated above hold for a base program
$\myprogram_B$ and its stable model $\smodel{B}$,
then a program $\myprogram_C$ obtained from $\myprogram_B$
by adding the translation of LTL formulas as given by Fig.~\ref{fig:trans},
has a stable model $\smodel{C}$, which follows the semantics of LTL for
all $0 \leq i \leq n$. 
\label{le:evalok}
\end{lemma}
\begin{proof}
First we note that Assumption \ref{ass:noocc}. above together with
Proposition~\ref{pro:strat:extension} and the fact that the translation
as given by Fig.~\ref{fig:trans} is stratified imply that 
a stable model $\smodel{C}$ of the combined program exists,
and is unique.

We now do the proof by induction on the structure of the formula $f$.
Assume that the translation of the subformulas $f_1$ and $f_2$
follow the semantics of LTL. Then we prove that also the translation
for $f$ follows the semantics of LTL.

We do a case split by the formula type:
\begin{itemize}
\item $f = p$, for $p \in \mathit{AP}$, or $f = \neg p$, for $p \in \mathit{AP}$:\\
By Assumption~\ref{ass:labelok}. above
$p(i) \in \smodel{C}$ iff $p(i) \in y_{(i)}$.
\item $f = f_1 \, \vee \, f_2$, or $f = f_1 \, \wedge \, f_2$:\\
The translation directly follows the semantics of LTL.
\item $f = f_1 \, U \, f_2$:
\\
We show that $f$ follows the semantics of LTL for all $0 \leq i \leq n$ 
by establishing that (i) it does that for $i=n$ and that (ii)
if $f$ follows the semantics of LTL for $i+1$, then it does for $i$. 
The proof is based on the following equivalence valid for the $U$
operator for all $0\leq i < n$: 
\begin{equation}
\mbox{$y^{(i)} \models f_1 \, U \, f_2$ iff 
$y^{(i)} \models f_2$ or 
($y^{(i)} \models f_1$ and $y^{(i+1)} \models f_1 \, U \, f_2$)}
\label{eq:Ufixedpoint}
\end{equation}

(i) Suppose $f(n) \in \smodel{C}$. In the cases b) and c) $f(n+1)
\not\in \smodel{C}$, implying that $f_2(n) \in \smodel{C}$ 
because $f(n) \lparrow f_2(n)$ is the only rule supporting $f(n)$. 
Hence, $y^{(n)} \models f$ holds and $y' \models f$ for any $y'$ extending
$y^{(n)}$ in the case c). Consider now the case a). Suppose
there is no $f_2(j) \in \smodel{C}$ with $l < j \leq n$. 
Then for $\smodel{C}' = \smodel{C} - \{f(j) \mid l < j \leq n\}$, 
$\smodel{C}' \models \reduct{\myprogram_C}{\smodel{C}}$ and 
$\smodel{C}' \subset \smodel{C}$ which implies that $\smodel{C}$ is not
a stable model of $\myprogram_C$, a contradiction. 
Hence, there is some $f_2(j) \in \smodel{C}$ with $l < j \leq n$. 
Take such $f_2(j) \in \smodel{C}$ with the smallest index $j$. 
Suppose there is some $f_1(j') \not\in \smodel{C}$ with $l < j' < j$.
Now for $\smodel{C}' = \smodel{C} - \{f(j')\}$, 
$\smodel{C}' \models \reduct{\myprogram_C}{\smodel{C}}$ and 
$\smodel{C}' \subset \smodel{C}$ implying that $\smodel{C}$ is not
a stable model of $\myprogram_C$, a contradiction. 
Hence, for all $l < j' < j$, $f_1(j) \in \smodel{C}$.
This implies by the inductive hypothesis that 
$y^{(n)} \models f$ holds.

Suppose $y^{(n)} \models f$ holds. Then in the case b) $y^{(n)} \models
f_2$ holds and hence, by rule $f(n) \lparrow f_2(n) \in \myprogram_C$,
$f(n) \in \smodel{C}$.
In the case a) there is some $f_2(j) \in \smodel{C}$ with $l < j \leq
n$, and for all $l < j' \leq j$, $f_1(j') \in \smodel{C}$.
Then by rules in the translation of $f$, $f(n) \in \smodel{C}$.

(ii) Suppose $f(i) \in \smodel{C}$, $i < n$. Then there is a supporting rule in
$\reduct{\myprogram_C}{\smodel{C}}$ with 
$f(i)$ in the head and the body literals satisfied in $\smodel{C}$. 
There are two candidate rules $f(i) \lparrow f_2(i)$ and 
$f(i) \lparrow f_1(i), \nexttime{f}{i}$. 
By the inductive hypotheses in the first case 
$y^{(i)} \models f_2$ and in the second case 
$y^{(i)} \models f_1$ and $y^{(i+1)} \models f$ which imply by
(\ref{eq:Ufixedpoint}) 
$y^{(i)} \models f$. 
In the other direction for cases a) and b), if 
$y^{(i)} \models f$, then by (\ref{eq:Ufixedpoint}) $y^{(i)} \models
f_2$ or ($y^{(i)} \models f_1$ and $y^{(i+1)} \models f$). From these
using the rules $f(i) \lparrow f_2(i)$ and 
$f(i) \lparrow f_1(i), \nexttime{f}{i}$ in $\smodel{C}$ and the 
inductive hypotheses follows that $f(i) \in \smodel{C}$ holds.

\item $f = f_1 \, R \, f_2$:\\
We show that $f$ follows the semantics of LTL for all $0 \leq i \leq n$ 
by establishing that (i) it does that for $i=n$ and that (ii)
if $f$ follows the semantics of LTL for $i+1$, then it does for $i$. 

(i) Consider first the case b). Now $f(n+1) \not\in \smodel{C}$. 
If $f(n) \in \smodel{C}$, then $f(n) \lparrow f_2(n)$ and $f(n) \lparrow
f_2(n),f_1(n)$ are the only rules
supporting $f(n)$ in $\reduct{\myprogram_C}{\smodel{C}}$. Hence, 
$f_2(n) \in \smodel{C}$, $y^{(n)} \models f_2$ and thus $y^{(n)} \models
f$. In the other direction, if  $y^{(n)} \models f$, then $y^{(n)} \models
f_2$ implying $f(n) \in \smodel{C}$. 
In the case c) if $f(n) \in \smodel{C}$, 
$f(n) \lparrow f_2(n),f_1(n)$ is the only rule
supporting $f(n)$ in $\reduct{\myprogram_C}{\smodel{C}}$
and hence $y^{(n)} \models f_1 \land f_2$ which implies 
$y' \models f$ for any $y'$ extending $y^{(n)}$.
Thus for the cases b) and c) condition (i) holds.

Now consider the case a) 
where we use the following equivalence between
LTL formulas:
\begin{equation}
\mbox{$y \models f_1 \, R \, f_2 $ iff $y \models (f_2 \, U \, (f_1 \, \wedge \, f_2 )) \, \vee \, (\Box f_2)$.}
\label{eq:Requiv}
\end{equation}
We consider two cases
\begin{itemize}
\item $y^{(n)} \models \Box f_2$

In this case by (\ref{eq:Requiv}) $y^{(n)} \models f$ but also 
$f_2(j) \in \smodel{C}$ for all  $l < j \leq n$ which
implies that $c(f) \not\in \smodel{C}$ and $f(n+1) \in \smodel{C}$ and
hence  $f(n) \in \smodel{C}$. 

\item $y^{(n)} \not\models \Box f_2$

In this case by (\ref{eq:Requiv}) 
$y^{(n)} \models f$ iff $y^{(n)} \models (f_2 \, U \, (f_1 \,
\wedge \, f_2 ))$. 
Now we can show that $y^{(n)} \models f$ iff $f(n) \in \smodel{C}$
using a similar argument as in the previous case for the $U$ operator.
This is because $y^{(n)} \not\models \Box f_2$ implies that 
there is some $f_2(j) \not\in \smodel{C}$ with  $l < j \leq
n$. Hence, $c(f) \in \smodel{C}$. Then the rules for $f(i)$ in 
$\reduct{\myprogram_C}{\smodel{C}}$ are 
\[
\begin{array}{l}
f(i) \lparrow f_2(i), f_1(i)\\
f(i) \lparrow f_2(i), \nexttime{f}{i} \\
\nexttrans{f} \lparrow \looppoint, f(i) \\
\end{array}
\]
which would be the evaluation
rules for the $U$ formula $(f_2 \, U \, (f_1 \, \wedge \, f_2 ))$. 
\end{itemize}
Hence, in both case $y^{(n)} \models f$ iff $f(n) \in \smodel{C}$ and
thus for the case a) condition (i) holds.

(ii) We use the following equivalence valid for all $0\leq i <n$: 
\begin{equation}
\mbox{$y^{(i)} \models f_1 \, R \, f_2$ iff 
$y^{(i)} \models f_2 \land f_1$ or 
($y^{(i)} \models f_2$ and $y^{(i+1)} \models f_1 \, R \, f_2$)}
\label{eq:Rfixedpoint}
\end{equation}
If $y^{(i)} \models f$, then by (\ref{eq:Rfixedpoint}) 
$f_2(i),f_1(i) \in
\smodel{C}$ or  $f_2(i),f(i+1) \in \smodel{C}$, which imply $f(i) \in
\smodel{C}$. 
In the other direction, if $f(i) \in\smodel{C}$, then there are two
possible rules in $\reduct{\myprogram_C}{\smodel{C}}$ supporting $f(i)$:
$f(i) \lparrow f_2(i), f_1(i)$ and
$f(i) \lparrow f_2(i), \nexttime{f}{i}$. Hence, 
$f_2(i),f_1(i) \in \smodel{C}$ or  $f_2(i),f(i+1) \in \smodel{C}$, which
imply by (\ref{eq:Rfixedpoint}) that $y^{(i)} \models f$. 
Hence, condition (ii) holds. 
\end{itemize}
\end{proof}

\paragraph{Final proof of Theorem~\ref{th:ltl}.}
Let $\smodel{LTL}$ be a stable of the program
$\myprogram = \mprogram{C_0}{0} \cup \aprogram{N}{n} \cup \lprogram{N}{n} \cup \ltlprogram{f}{n}$.
By Lemma~\ref{le:maxex} we can obtain from $\smodel{LTL}$
a maximal (interleaving) execution
$\sigma$ of $N$ from an initial marking satisfying $C_0$.
Consider now the subprogram $\myprogram_{B}$, which consist
of $\mprogram{C_0}{0} \cup \aprogram{N}{n} \cup \lprogram{N}{n}$
and the rules (\ref{rule:guessloop})-(\ref{rule:onevisible})
of $\ltlprogram{f}{n}$. By Proposition~\ref{pro:projection}
and Lemma~\ref{le:compatible} the stable model
$\smodel{LTL}$ projected on the atoms of $\myprogram_{B}$
satisfies the assumptions required by Lemma~\ref{le:evalok}.
Now Lemma~\ref{le:maxstepex} and
Lemma~\ref{le:evalok} with $u = w'$
imply that if $f(0)$ is in a stable model $\smodel{LTL}$,
then for the word $w''$ corresponding to the maximal step execution $\sigma''$
it holds that $w'' \models f$.
Because the word $w''$
is stuttering equivalent to the word $w$ corresponding to
$\sigma$ according to Lemma~\ref{le:st:to:inter},
Proposition~\ref{pro:stutt} implies that
if $f(0)$ is in a stable model $\smodel{LTL}$,
then $\sigma \models f$.
The rule (\ref{eq:topok}) implies that $f(0) \in \smodel{LTL}$,
and thus $\sigma \models f$.
This completes our proof of Theorem~\ref{th:ltl}.
\proofbox

\subsection{Proof of Theorem \ref{th:compl}\label{app:compl}}

We first recall our proof objective.
Let $f$ be an LTL formula in positive normal form and
$N = \seq{P, T, F}$ be a 1-safe P/T-net for all initial markings
satisfying a condition $C_0$.

We want to prove that if $N$ has a looping or deadlock execution of at most length $n$
starting from an initial marking satisfying $C_0$ such that some corresponding
maximal execution $\sigma$ satisfies $f$, then
$\myprogram = \mprogram{C_0}{0} \cup \aprogram{N}{n} \cup \lprogram{N}{n} \cup \ltlprogram{f}{n}$
has a stable model.

We thus know that there is a deadlock or looping execution,
call it $\sigma'$, of $N$ of length $n'$ such that
$n' \leq n$ and for some corresponding maximal execution $\sigma$
it holds that $\sigma \models f$.

There are now two mutually exclusive cases:
\begin{itemize}
\item[a)] $\sigma'$ is a looping execution.
Notice that in this case $S_{n'-1} = \{ t_{n'-1} \}$, i.e., the last step is always
non-empty.
Without loss of generality
we select the minimal index $0 \leq l \leq n-1$ such that $M_{l} = M_{n}$ and
such that the corresponding maximal execution $\sigma \models f$,
where $\sigma$ is the maximal execution which visits
the sequence of states $M_0, M_1, \ldots, M_l, M_{l+1}, \ldots, M_k, M_{l+1}, \ldots, M_k, \ldots$.
\item[b)] $\sigma'$ is a deadlock execution. In this case $\sigma = \sigma'$ is
a maximal execution such that $\sigma \models f$.
We now set $l = n$ to differentiate from the previous case.
\end{itemize}

Note that the case c) used in proof of Theorem~\ref{th:ltl} is not
needed here, as we do not consider non-maximal executions.

Lemma~\ref{le:execution:to:model} implies that
the program $\myprogram_0 = \mprogram{C_0}{0} \cup \aprogram{N}{n}$ has a stable
model $\smodel{0}$ whose derived execution can be obtained from
$\sigma'$ by adding $n - n'$ empty steps at the beginning of the execution.

We keep the name $\sigma'$ for the step execution derived
by adding $n - n'$ idle steps to the beginning of $\sigma'$,
and add $n - n'$ idle steps to the beginning of the
corresponding maximal execution $\sigma$.
Also the loop point $l$ is increased by $n - n'$
to compensate for the addition of idle steps.
Clearly the obtained step execution $\sigma'$ is of length $n$,
is a deadlock execution iff the original is, and
is a looping execution iff the original is.
Moreover, the corresponding maximal execution $\sigma$ satisfies the
formula $f$ also after the addition of the idle steps,
as the word $w$ corresponding to $\sigma$ has
$n -n'$ extra copies of stuttering of the initial
atomic propositions to it, which by Proposition~\ref{pro:stutt}
cannot be detected by any LTL formula.
Therefore we can from now on assume that $\sigma'$ is of length
(exactly) $n$.

Now consider the program
$\myprogram_2$ which consists of
$\mprogram{C_0}{0} \cup \aprogram{N}{n} \cup \lprogram{N}{n}$
together with rules (\ref{rule:guessloop})-(\ref{rule:onevisible})
of program $\ltlprogram{f}{n}$.
Given the pair $(\sigma',l)$ we will show that the
program $\myprogram_2$ has a stable model $\smodel{2}$
capturing the essential properties of $(\sigma', l)$.

\begin{lemma}
Given the pair $(\sigma',l)$ we can construct a stable model
$\smodel{2}$ of the program $\myprogram_{2}$ such that the
two claims stated below hold for $\smodel{2}$.
\begin{enumerate}
\item For all $p \in P$, $0 \leq i \leq n$: $p(i) \in \smodel{2}$
iff $p(i) \in M_{i}$ in $\sigma'$.
\item $\smodel{2}$ is exactly one of the
following two cases:
\begin{itemize}
\item[a)] If $0 \leq l \leq n-1$, then 
$\mathit{nl}(l+1) \in \smodel{2}$, $\mathit{nl}(i) \not \in \smodel{2}$
for all $i \not = (l + 1)$,
$\mathit{le} \in \smodel{2}$,
$\mathit{il}(i) \in \smodel{2}$ for all $l+1 \leq i \leq n$, and
$\mathit{live} \in \smodel{2}$.
\item[b)] If $l = n$, then
$\mathit{nl}(i+1) \not \in \smodel{2}$ for all $0 \leq i \leq n-1$,
$\mathit{le} \not \in \smodel{2}$, and
$\mathit{live} \not \in \smodel{2}$.
\end{itemize}
\end{enumerate}
\label{le:assok2}
\end{lemma}
\begin{proof}
In the proofs below, use case a) when $0 \leq l \leq n-1$, and the case
b) when $l = n$.

We know that the first claim holds for the stable model
$\smodel{0}$ of the program $\myprogram_{0}$ as
$\sigma'$ has been derived from it.

We also know from Lemma~\ref{le:livelockok}, and
the fact that $\sigma'$ has been derived from the stable model 
$\smodel{0}$ of $\myprogram_{0}$
that the program
$\myprogram_1 = \myprogram_{0} \cup \lprogram{N}{n}$
has a stable model $\smodel{1}$, such that
\begin{itemize}
\item[a)] $\smodel{1} = \smodel{0} \cup \{ \mathit{live} \}$,
as a loop execution is not deadlocked after $\sigma'$, or
\item[b)] $\smodel{1} = \smodel{0}$,
as $\sigma'$ is a deadlock execution.
\end{itemize}

Given $\myprogram_{1}$ and the stable model $\smodel{1}$,
we will incrementally add rules to the program $\myprogram_1$
whose head atoms do not appear in the program they are added into.
At each step we prove that a stable model of the extended program
exists. At the end we use Proposition~\ref{pro:projection} to project
the final stable model $\smodel{2}$ on the atoms of $\myprogram_{1}$ obtaining
$\smodel{1}$, which fulfils the first claim and part of the second claim.

The rest of the second claim is proved incrementally by stating properties
the stable models extending $\smodel{1}$ will have, finally
ending up with the stable model $\smodel{2}$ of $\myprogram_{2}$
which satisfies also the rest of the second claim.

\begin{enumerate}
\item First add the shorthand rules (\ref{rule:guessloop}) to $\myprogram_{1}$
obtaining the program $\myprogram_{a}$. 
We show that a stable model $\smodel{a}$ exists, which extends
$\smodel{1}$ as follows. We do case analysis:
\begin{itemize}
\item[a)] $\smodel{a} = \smodel{1} \cup \{ \mathit{el(l)}\} \cup
\{ el(i)' \, \arrowvert \, 0 \leq i \leq n-1 \mbox{ such that } i \not = l\}$
\item[b)] $\smodel{a} = \smodel{1} \cup \{ el(i)' \, \arrowvert \, 0 \leq i \leq n-1 \}$
\end{itemize}
In the case a) the shorthands (\ref{rule:guessloop}) contribute to the
reduct $\reduct{\myprogram_a}{\smodel{a}}$ a fact $el(l) \lparrow$ and
a fact $el(i)' \lparrow$ for every $0 \leq i \leq n-1$ such that $i \not=
l$. Clearly, $\smodel{a} \models \reduct{\myprogram_a}{\smodel{a}}$
and, moreover, $\smodel{a}$ is the smallest such set $\smodel{} \subseteq
\smodel{a}$ because removing  $el(l)$ or one of $el(i)'$ would leave the
corresponding fact unsatisfied. Hence, $\smodel{a}$ is a stable model of
$\myprogram_a$. The case b) is similar except that there is no fact 
$el(l) \lparrow$ in the reduct but a fact $el(i)' \lparrow$ for every $0
\leq i \leq n-1$.

\item Add the integrity constraints (\ref{rule:guesscheck})-(\ref{rule:checkloop})
to $\myprogram_{a}$
obtaining $\myprogram_{b}$. Now
$\smodel{b} = \smodel{a}$ is a stable model of $\myprogram_{b}$.
In the case a) the integrity constraint (\ref{rule:guesscheck}) is satisfied
because $el(i) \in \smodel{b}$ only for the index $i = l$.
Because $\sigma'$ is a loop execution with $M_l = M_k$ also all of the
integrity constraints (\ref{rule:checkloop}) are satisfied.
In the case b) $el(i) \not \in \smodel{b}$ for all indices $0 \leq i \leq n-1$,
and thus the integrity constraint (\ref{rule:guesscheck}) is satisfied
and also all of the integrity constraints (\ref{rule:checkloop}) are satisfied.

\item Add rules (\ref{rule:loopstuff}) to $\myprogram_{b}$ obtaining $\myprogram_{c}$.
The added rules are stratified, and
thus by Proposition~\ref{pro:strat:extension}
a~unique stable model $\smodel{c}$ exists, which extends $\smodel{b}$ as follows.
We do a case analysis:
\begin{itemize}
\item[a)] $\smodel{c} = \smodel{b} \cup \{ \mathit{le}, \mathit{nl}(l+1) \} \cup
\{ \mathit{il}(i) \, \arrowvert \, l+1 \leq i \leq n \}$
\item[b)] $\smodel{c} = \smodel{b}$
\end{itemize}
The proof in the case a) proceeds starting from the fact that $el(l) \in \smodel{b}$,
from which we get $\{ \mathit{le}, \mathit{nl}(l+1), \mathit{il}(l+1)\} \in \smodel{c}$.
We then get that $\{ \mathit{il}(i) \, \arrowvert \, l+2 \leq i \leq n \} \in \smodel{c}$
by simple induction using the rules $\mathit{il}(i+1) \lparrow \mathit{il}(i)$.
In the case b) $\mathit{el}(i) \not \in \smodel{b}$ for all indices $0 \leq i \leq n-1$ implies
that the rules (\ref{rule:loopstuff}) are satisfied by $\smodel{c}$,
and thus $\smodel{c}$ is a stable model.

\item Add the rule~(\ref{rule:noidleloop}) to $\myprogram_{c}$
obtaining $\myprogram_{d}$.
This is an integrity constraint, which is satisfied
in case a), as the step $S_{n-1}$ is non-empty in looping executions.
The integrity constraint is also satisfied in case b), as $\mathit{le} \not \in \smodel{c}$.
Thus $\smodel{d} = \smodel{c}$ is in both cases a stable model of $\myprogram_{d}$.

\item Finally, add the integrity constraints (\ref{rule:onevisible}) to $\myprogram_{d}$ obtaining
$\myprogram_{e}$. These integrity constraints are always satisfied, as no time step
in $\sigma'$ contains more than one transition.
Thus $\smodel{e} = \smodel{d}$ will be a stable model of $\myprogram_{e}$.
\end{enumerate}

By setting $\smodel{2} = \smodel{e}$ we have shown that $\smodel{2}$ is a stable
model of the program $\myprogram_2 = \myprogram_{e}$ such that $\smodel{2}$ projected on the atoms
of $\myprogram_1$ by Proposition~\ref{pro:projection} is
$\smodel{1}$. This satisfies the first claim, and the part of the second claim
concerning the atom $\mathit{live}$.
The rest of the second claim has been incrementally proved in the cases above.
\end{proof}

Let $\myprogram_3$ be the program which consists of
$\myprogram_2$ together with the translation of
the LTL formula $f$ as given by Fig.~\ref{fig:trans},
and let $w'$ be the finite word corresponding to the execution $\sigma'$.

Lemma~\ref{le:assok2} and Lemma~\ref{le:evalok}
with $u = w'$ and $\myprogram_{B} = \myprogram_2$, and the fact that
$\sigma \models f$ implies that the program $\myprogram_3$ has a stable model
such that $f(0) \in \smodel{3}$.
(Recall that we do not use case c), and thus our evaluation of $f$ on
the corresponding maximal execution $\sigma$ is exact.)
We can now add the constraint rule (\ref{eq:topok}) to the program
$\myprogram_3$, and to obtain the full program $\myprogram$.
Now because $f(0) \in \smodel{3}$ the integrity constraint (\ref{eq:topok})
is satisfied, and we have found a stable model
$\smodel{} = \smodel{3}$ of $\myprogram$.
This completes our proof of Theorem~\ref{th:compl}.
\proofbox

\end{document}